# Proceedings Scholar Metrics:
# H Index of proceedings on Computer Science, Electrical & Electronic Engineering, and Communications according to Google Scholar Metrics (2009-2013)

**Alberto Martín-Martín[1], Enrique Orduña-Malea[2], Juan Manuel Ayllón[1], Emilio Delgado López-Cózar[1]**

[1]EC3: Evaluación de la Ciencia y de la Comunicación Científica. Universidad de Granada
[2]EC3: Evaluación de la Ciencia y de la Comunicación Científica, Universidad Politécnica de Valencia (Spain)

## ABSTRACT

The objective of this report is to present a list of proceedings (conferences, workshops, symposia, meetings) in the areas of Computer Science, Electrical & Electronic Engineering, and Communications covered by Google Scholar Metrics and ranked according to their h-index. Google Scholar Metrics only displays publications that have published at least 100 papers and have received at least one citation in the last five years (2009-2013). The searches were conducted between the 15th and 22nd of December, 2014. A total of 1208 proceedings have been identified.

**Corresponding authors:**
**Emilio Delgado López-Cózar. edelgado@ugr.es**

# INTRODUCTION

**PROCEEDINGS SCHOLAR METRICS** is a ranking that displays proceedings (conferences, workshops, symposia, meetings) indexed in Google Scholar Metrics (GSM) on the areas of Computer Science, Electrical & Electronic Engineering, and Communications for the period 2009-2013.

It is a well-known fact that conference proceedings play a major role as a means of scientific communication in all areas concerning Computer Engineering, Electronics, and Communications. The rapid rate at which knowledge is generated in these fields required the creation of a more dynamic system to communicate and publish research results. Conferences have historically fulfilled this role.

Therefore, it is not surprising that these publications take up an important place in researchers' curriculum and that it is an object of special consideration in researchers' performance evaluation programmes.

The various databases that have been traditionally used for evaluative purposes (Web of Science, Scopus) cover these types of publications with varying degrees of exhaustiveness, but never optimally. In fact, this has been one of the reasons why there have already been various attempts at creating classifications that identify and rank these types of publications.

The creation of Google Scholar in 2004, an academic search engine dedicated to trawling the Web in search of scientific literature, revolutionized the world of scientific information search systems. Particularly, it had very positive effect on those disciplines where publication habits were not limited to publishing in scientific journals, i.e. Engineering, Computer Science, Electrical & Electronic Engineering, and Communications.

Publications in these disciplines, for which the Internet is the natural environment, are splendidly represented on the Web. That is why Google Scholar is especially useful in these fields.

The development of Google Scholar Metrics, launched on April 2012 with the goal of providing a ranking of scientific publications indexed on Google Scholar (journals, proceedings, repositories), provided that they had published at least 100 papers and received at least one citation in the last five years, has been a crucial step towards knowing the impact of conferences, which are so important in these areas.

In GSM, publications rankings are sorted by impact (h index) and they can be browsed by languages, each of these showing the 100 publications with the greatest impact. For publications written in English, this tool allows user to browse publications by subject categories and subcategories. Computer Science, Engineering and Communications can be found on the "Engineering & Computer Science Category", which comprises 58 subcategories. For each of these subcategories, only the top 20 publications with a highest H Index are displayed. There are a total of 190 unique conference proceedings across these subcategories. Categories and subcategories are not available for the other eight languages present in GSM (Chinese, Portuguese, German, Spanish, French, Italian, Japanese and Duth). In order to find other publications not displayed in the language and category league lists, there is a search box at our disposal. However, any given query will only display 20 results.

Therefore, the system interface does not allow us to effectively determine which and how many conference proceedings GSM has indexed. In order to overcome this limitation, the objective of PROCEEDINGS SCHOLAR METRICS is to compile an inventory of all the conferences present in GSM concerning these fields of knowledge, and after that, rank them according to their scientific impact, as measured by the H index.

# MATERIAL AND METHODS

### Subject areas covered

Proceedings concerning Computer Science (theoretical, information theory, artificial intelligence, evolutionary computation, fuzzy systems, human computer interaction, computer vision & pattern recognition, computer hardware design, computing systems, signal processing, computer networks & wireless communication, robotics, automation & control theory, software systems, computer security & cryptography, computer graphics, databases & information systems, data mining & analysis, multimedia, bioinformatics & computational biology, biomedical technology, medical informatics, computational linguistics, education technology), Electrical & Electronic Engineering and Communications (telecommunications, remote sensing, antennas, radar, microware).

### Search strategy

In order to identify the proceedings we followed two different strategies:

1. We collected all the conferences displayed in the subcategories concerning Computer Science, Electrical & Electronic Engineering, and Communications.
2. We carried out various searches using descriptive and pertinent keywords in order to locate the rest of relevant conferences. These searches took place on the third week of December, 2014.

The results of each search were downloaded (including the name of the conference, the h5-index, and the h5-median) and duplicates were removed. A manual check was carried out in order to filter out any irrelevant entries (journals, repositories, and conferences outside the scope of our study). A total of 1208 conference proceedings were identified.

### Criteria for the inclusion of Google Scholar Metrics proceedings

Proceedings that published at least 100 papers and received at least one citation in the last five years (2009-2013)

### Sorting criteria and fields displayed

The proceedings are sorted by their H Index. In case of a tie, the discriminate value is the h5-median (the median number of citations for the articles that make up its h-index)

The information displayed for each conference is:
- H5-Index
- H5-median
- Quartile: quartile position of the conference.
- Thematic subcategories the conference has been assigned to.

# H Index of proceedings on Computer Science, Electrical & Electronic Engineering, and Communications according to Google Scholar Metrics (2009-2013)

| Rank | Quartil | Proceedings | h-index | h5-median |
|---|---|---|---|---|
| 1 | Q1 | IEEE Conference on Computer Vision and Pattern Recognition, CVPR | 118 | 167 |
| 2 | Q1 | Proceedings of the IEEE | 85 | 162 |
| 3 | Q1 | International World Wide Web Conferences (WWW) | 83 | 131 |
| 4 | Q1 | IEEE International Conference on Computer Vision, ICCV | 79 | 138 |
| 5 | Q1 | Annual Joint Conference of the IEEE Computer and Communications Societies (INFOCOM) | 76 | 110 |
| 6 | Q1 | ACM SIGKDD International Conference on Knowledge discovery and data mining | 69 | 113 |
| 7 | Q1 | International Conference on Machine Learning (ICML) | 69 | 103 |
| 8 | Q1 | ACM SIGCOMM Conference | 67 | 131 |
| 9 | Q1 | Neural Information Processing Systems (NIPS) | 66 | 94 |
| 10 | Q1 | ACM Symposium on Information, Computer and Communications Security | 65 | 110 |
| 11 | Q1 | International Conference on Very Large Databases | 64 | 92 |
| 12 | Q1 | Meeting of the Association for Computational Linguistics (ACL) | 62 | 84 |
| 13 | Q1 | ACM SIGMOD Conference | 61 | 90 |
| 14 | Q1 | IEEE International Conference on Robotics and Automation | 61 | 87 |
| 15 | Q1 | European Conference on Computer Vision | 59 | 96 |
| 16 | Q1 | IEEE International Solid-State Circuits Conference | 59 | 79 |
| 17 | Q1 | International Conference on Software Engineering | 57 | 94 |
| 18 | Q1 | Symposium on Networked Systems: Design and Implementation (NSDI) | 54 | 84 |
| 19 | Q1 | ACM International Conference on Web Search and Data Mining | 54 | 82 |
| 20 | Q1 | International Symposium on Computer Architecture (ISCA) | 53 | 98 |
| 21 | Q1 | IEEE Symposium on Security and Privacy | 53 | 85 |
| 22 | Q1 | Conference on Empirical Methods in Natural Language Processing (EMNLP) | 53 | 78 |
| 23 | Q1 | International Conference on Data Engineering Workshops | 52 | 68 |
| 24 | Q1 | International Conference on Architectural Support for Programming Languages and Operating Systems (ASPLOS) | 51 | 87 |
| 25 | Q1 | ACM Symposium on Theory of Computing | 51 | 77 |
| 26 | Q1 | USENIX Security Symposium | 51 | 76 |
| 27 | Q1 | IEEE Transactions on Microwave Theory and Techniques | 51 | 61 |
| 28 | Q1 | IEEE International Symposium on Information Theory | 49 | 80 |
| 29 | Q1 | ACM SIGIR Conference on Research and development in information retrieval | 49 | 71 |
| 30 | Q1 | International Conference on Weblogs and Social Media | 48 | 87 |
| 31 | Q1 | Annual International Conference on Theory and Applications of Cryptographic Techniques (EUROCRYPT) | 48 | 79 |
| 32 | Q1 | ACM International Conference on Information and Knowledge Management | 48 | 62 |
| 33 | Q1 | International Conference on Mobile systems, applications, and services | 47 | 94 |
| 34 | Q1 | Internet Measurement Conference | 47 | 78 |
| 35 | Q1 | IEEE International Conference on Acoustics, Speech and Signal Processing (ICASSP) | 47 | 65 |
| 36 | Q1 | Annual International Conference on Mobile computing and networking | 46 | 72 |
| 37 | Q1 | SIGPLAN Conference on Programming Language Design and Implementation (PLDI) | 46 | 71 |
| 38 | Q1 | IEEE International Symposium on Parallel & Distributed Processing | 46 | 67 |
| 39 | Q1 | International Joint Conference on Artificial Intelligence (IJCAI) | 46 | 63 |
| 40 | Q1 | Conference on Advances in cryptology | 45 | 77 |
| 41 | Q1 | IEEE GLOBECOM Workshops | 45 | 63 |
| 42 | Q1 | IEEE/RSJ International Conference on Intelligent Robots and Systems | 45 | 62 |
| 43 | Q1 | ACM SIGPLAN-SIGACT Symposium on Principles of Programming Languages (POPL) | 45 | 61 |
| 44 | Q1 | IEEE Symposium on Foundations of Computer Science (FOCS) | 45 | 58 |
| 45 | Q1 | IEEE International Conference on Communications | 44 | 62 |
| 46 | Q1 | ACM International Conference on Multimedia | 44 | 60 |
| 47 | Q1 | IEEE Conference on Decision and Control | 43 | 63 |
| 48 | Q1 | ACM SIAM Symposium on Discrete Algorithms | 43 | 59 |
| 49 | Q1 | ACM European Conference on Computer Systems | 42 | 81 |
| 50 | Q1 | IEEE International Conference on Cloud Computing (CLOUD) | 42 | 62 |
| 50 | Q1 | IEEE International Symposium on High Performance Computer Architecture | 42 | 62 |
| 52 | Q1 | IEEE Vehicular Technology Conference, VTC | 42 | 59 |
| 53 | Q1 | International Conference on Autonomous Agents and Multiagent Systems | 42 | 54 |
| 54 | Q1 | Design, Automation and Test in Europe Conference and Exhibition (DATE) | 41 | 57 |
| 55 | Q1 | ACM Symposium on User Interface Software and Technology | 40 | 62 |
| 55 | Q1 | IEEE/ACM International Symposium on Microarchitecture | 40 | 62 |
| 57 | Q1 | Design Automation Conference (DAC) | 40 | 54 |
| 58 | Q1 | Network and Distributed System Security Symposium (NDSS) | 39 | 79 |
| 59 | Q1 | International Conference on Spoken Language Processing (INTERSPEECH) | 39 | 52 |
| 60 | Q1 | IEEE Microwave and Wireless Components Letters | 39 | 50 |
| 61 | Q1 | American Control Conference | 39 | 48 |
| 62 | Q1 | IEEE Applied Power Electronics Conference and Exposition | 38 | 61 |
| 63 | Q1 | International Conference on Computer Aided Verification (CAV) | 38 | 53 |
| 64 | Q1 | Hawaii International Conference on System Sciences | 38 | 49 |
| 65 | Q1 | European Semantic Web Symposium / Conference | 38 | 48 |
| 66 | Q1 | Quantum Electronics and Laser Science Conference | 37 | 70 |
| 67 | Q1 | Conference on File and Storage Technologies (FAST) | 37 | 62 |
| 68 | Q1 | Conference on Computer Supported Cooperative Work (CSCW) | 37 | 60 |
| 69 | Q1 | IEEE International Conference on Cloud Computing Technology and Science (CloudCom) | 37 | 58 |
| 70 | Q1 | British Machine Vision Conference (BMVC) | 37 | 56 |
| 71 | Q1 | AAAI Conference on Artificial Intelligence | 37 | 50 |
| 72 | Q1 | ACM Conference on Recommender Systems | 36 | 62 |
| 73 | Q1 | International Conference on Distributed Computing Systems, ICDCS | 36 | 49 |
| 73 | Q1 | International Symposium on Information Processing in Sensor Networks | 36 | 49 |
| 75 | Q1 | IEEE International Conference on Data Mining (ICDM) | 36 | 47 |
| 75 | Q1 | International Conference on Extending Database Technology (EDBT) | 36 | 47 |
| 77 | Q1 | ACM Conference on Embedded Networked Sensor Systems | 35 | 81 |
| 78 | Q1 | ACM SIGPLAN Symposium on Principles & Practice of Parallel Programming (PPOPP) | 35 | 69 |
| 79 | Q1 | USENIX Annual Technical Conference | 35 | 54 |
| 80 | Q1 | SIAM International Conference on Data Mining | 35 | 49 |
| 81 | Q1 | IEEE International Conference on Image Processing (ICIP) | 35 | 47 |
| 82 | Q1 | International Conference on The Theory and Application of Cryptology and Information Security (ASIACRYPT) | 34 | 58 |
| 83 | Q1 | Conference on Object-Oriented Programming Systems, Languages, and Applications (OOPSLA) | 34 | 50 |
| 83 | Q1 | IEEE International Symposium on Cluster Computing and the Grid | 34 | 50 |
| 85 | Q1 | IEEE International Conference on Web Services | 34 | 46 |
| 86 | Q1 | Annual Symposium on Cloud Computing | 33 | 64 |
| 87 | Q1 | IEEE International Conference on Computer Vision Workshops (ICCV Workshops) | 33 | 59 |
| 88 | Q1 | Allerton Conference on Communication, Control, and Computing | 33 | 49 |
| 89 | Q1 | Workshop on Cryptographic Hardware and Embedded Systems (CHES) | 33 | 45 |
| 90 | Q1 | International Quantum Electronics Conference | 32 | 57 |
| 91 | Q1 | International Conference on Parallel Architectures and Compilation Techniques (PACT) | 32 | 52 |
| 92 | Q1 | Conference in Uncertainty in Artificial Intelligence | 32 | 49 |
| 93 | Q1 | ACM Conference on Electronic Commerce | 32 | 47 |
| 94 | Q1 | IEEE International Conference on Pervasive Computing and Communications Workshops (PERCOM Workshops) | 32 | 45 |
| 95 | Q1 | ACM Symposium on Applied Computing | 32 | 39 |
| 95 | Q1 | IEEE Congress on Evolutionary Computation | 32 | 39 |
| 97 | Q1 | Medical Image Computing and Computer Assisted Intervention | 32 | 37 |
| 98 | Q1 | Conference on Emerging Network Experiment and Technology | 31 | 63 |
| 99 | Q1 | ACM International Symposium on Mobile Ad Hoc Networking and Computing | 31 | 52 |
| 100 | Q1 | International Conference on Computational Linguistics (COLING) | 31 | 48 |
| 101 | Q1 | International Symposium on Software Testing and Analysis | 31 | 47 |
| 102 | Q1 | International Conference on Tools and Algorithms for the Construction and Analysis of Systems (TACAS) | 31 | 44 |
| 103 | Q1 | International Symposium/Conference on Music Information Retrieval | 31 | 43 |
| 104 | Q1 | IEEE Wireless Communications & Networking Conference | 31 | 42 |
| 105 | Q1 | International Conference on Document Analysis and Recognition | 31 | 41 |
| 106 | Q1 | Conference on Genetic and Evolutionary Computation | 31 | 39 |
| 107 | Q1 | IEEE Real-Time Systems Symposium (RTSS) | 30 | 53 |
| 107 | Q1 | International Conference on Practice and Theory in Public Key Cryptography | 30 | 53 |
| 109 | Q1 | International Conference on Supercomputing (ICS) | 30 | 50 |
| 110 | Q1 | ACM International Symposium on High Performance Distributed Computing | 30 | 48 |
| 111 | Q1 | Conference of the European Chapter of the Association for Computational Linguistics (EACL) | 30 | 45 |
| 111 | Q1 | IEEE Microwave Magazine | 30 | 45 |
| 113 | Q1 | IEEE/ACM International Conference on Automated Software Engineering (ASE) | 30 | 44 |
| 114 | Q1 | International Conference on Advanced Information Systems Engineering | 30 | 43 |
| 115 | Q1 | IEEE International Conference on Software Testing, Verification and Validation Workshops (ICSTW) | 30 | 38 |
| 116 | Q1 | Computer Security Applications Conference | 29 | 50 |
| 117 | Q1 | ACM SIGSPATIAL International Conference on Advances in Geographic Information Systems | 29 | 47 |
| 117 | Q1 | Symposium On Usable Privacy and Security | 29 | 47 |
| 119 | Q1 | ACM/IEEE International Conference on Human Robot Interaction | 29 | 45 |
| 120 | Q1 | ACM Symposium on Parallelism in Algorithms and Architectures (SPAA) | 29 | 44 |
| 121 | Q1 | IEEE Conference on Computer Communications Workshops | 29 | 43 |
| 121 | Q1 | IEEE International Conference on Communications Workshops | 29 | 43 |
| 123 | Q1 | Business Process Management | 29 | 40 |
| 124 | Q1 | IEEE/ACM International Conference on Computer-Aided Design (ICCAD) | 29 | 39 |
| 125 | Q1 | Symposium on Theoretical Aspects of Computer Science (STACS) | 29 | 38 |
| 126 | Q1 | IFAC World Congress | 29 | 35 |
| 126 | Q1 | International Conference on Pattern Recognition | 29 | 35 |
| 128 | Q1 | CIDR [Conference on Innovative Data Systems Research] | 28 | 61 |
| 129 | Q1 | European Conference on Research in Computer Security | 28 | 49 |
| 130 | Q1 | European Conference on Object-oriented Programming (ECOOP) | 28 | 45 |
| 131 | Q1 | European Conference on Advances in Information Retrieval | 28 | 43 |
| 132 | Q1 | International Conference on Functional Programming (ICFP) | 28 | 42 |
| 133 | Q1 | International Conference on Financial Cryptography and Data Security | 28 | 40 |
| 134 | Q1 | ACM SIGMOD-SIGACT-SIGART Symposium on Principles of Database Systems | 28 | 39 |
| 135 | Q1 | Artificial Intelligence in Medicine | 28 | 38 |
| 136 | Q1 | IEEE/IFIP International Conference on Dependable Systems and Networks | 28 | 37 |
| 137 | Q1 | Asia and South Pacific Design Automation Conference (ASP-DAC) | 28 | 36 |
| 137 | Q1 | IEEE Information Theory Workshop | 28 | 36 |
| 139 | Q1 | European Conference on Machine learning and knowledge discovery in databases | 27 | 40 |
| 139 | Q1 | IEEE Computer Security Foundations Symposium | 27 | 40 |
| 139 | Q1 | IEEE International Symposium on Personal, Indoor and Mobile Radio Communications | 27 | 40 |
| 142 | Q1 | ACM International Conference on Interactive Tabletops and Surfaces (ITS) | 27 | 39 |
| 142 | Q1 | IEEE Conference of Industrial Electronics | 27 | 39 |
| 144 | Q1 | ACM Symposium on Principles of Distributed Computing (PODC) | 27 | 37 |
| 145 | Q1 | International Conference on Biometrics | 27 | 34 |
| 146 | Q1 | IEEE International Symposium on Circuits and Systems | 27 | 33 |
| 147 | Q1 | IEEE/MTT-S International Microwave Symposium | 27 | 32 |
| 148 | Q1 | Conference on Information Sciences and Systems | 26 | 38 |
| 148 | Q1 | International Conference on Intelligent User Interfaces (IUI) | 26 | 38 |
| 148 | Q1 | Symposium on Interactive 3D Graphics (SI3D) | 26 | 38 |
| 151 | Q1 | Symposium on Field Programmable Gate Arrays (FPGA) | 26 | 37 |
| 152 | Q1 | ACM SIGGRAPH/Eurographics Symposium on Computer Animation | 26 | 36 |
| 152 | Q1 | IEEE Symposium on Logic in Computer Science | 26 | 36 |
| 152 | Q1 | International Symposium on Code Generation and Optimization | 26 | 36 |
| 155 | Q1 | Electronic Components and Technology Conference, ECTC | 26 | 35 |
| 155 | Q1 | International Joint Conference on Neural Networks | 26 | 35 |
| 155 | Q1 | Microwave and Optical Technology Letters | 26 | 35 |
| 158 | Q1 | European Conference on Parallel Processing (Euro-Par) | 26 | 33 |
| 159 | Q1 | ACM Workshop on Hot Topics in Networks | 25 | 51 |
| 160 | Q1 | IEEE International Conference on Cluster Computing | 25 | 47 |
| 161 | Q1 | International Journal of Computer-Supported Collaborative Learning | 25 | 43 |
| 162 | Q1 | European Symposium on Programming (ESOP) | 25 | 42 |
| 163 | Q1 | IEEE International Conference on Software Maintenance | 25 | 40 |
| 164 | Q1 | European Conference and Exposition on Optical Communications | 25 | 38 |
| 165 | Q1 | Asian Conference on Computer Vision | 25 | 36 |
| 166 | Q1 | European Conference on Algorithms | 25 | 35 |
| 166 | Q1 | Symposium on Computational Geometry | 25 | 35 |
| 168 | Q1 | IEEE Conference on Computational Intelligence and Games | 25 | 34 |
| 169 | Q1 | IEEE International Conference on Services Computing | 25 | 33 |
| 170 | Q1 | IEEE Consumer Communications and Networking Conference | 25 | 32 |
| 171 | Q1 | International Conference on Information Systems (ICIS) | 25 | 30 |
| 172 | Q1 | International Conference on Automated Planning and Scheduling (ICAPS) | 25 | 23 |
| 173 | Q1 | European Quantum Electronics Conference | 24 | 50 |
| 174 | Q1 | International Software Product Line Conference | 24 | 47 |
| 175 | Q1 | International Conference on Autonomic computing | 24 | 40 |
| 176 | Q1 | IET Microwaves, Antennas & Propagation | 24 | 39 |
| 176 | Q1 | Information Theory and Applications Workshop | 24 | 39 |
| 176 | Q1 | Workshop of Cross-Language Evaluation Forum | 24 | 39 |
| 179 | Q1 | ACM/IEEE International Symposium on Low Power Electronics and Design | 24 | 38 |
| 180 | Q1 | European Conference on Optical Communications | 24 | 36 |
| 180 | Q1 | International Conference on Affective Computing and Intelligent Interaction and Workshops | 24 | 36 |
| 182 | Q1 | ACM/IEEE International Symposium on Networks-on-Chip | 24 | 35 |
| 183 | Q1 | Human-Computer Interaction | 24 | 34 |
| 184 | Q1 | European Conference on Power Electronics and Applications | 24 | 31 |
| 184 | Q1 | International Conference on Tangible, embedded, and embodied interaction | 24 | 31 |
| 184 | Q1 | International Symposium on Empirical Software Engineering and Measurement, ESEM | 24 | 31 |
| 184 | Q1 | IEEE Real-Time and Embedded Technology and Applications Symposium | 24 | 31 |
| 188 | Q1 | IFIP Conference on Human-Computer Interaction (INTERACT) | 24 | 28 |
| 189 | Q1 | IEEE International Symposium on Mixed and Augmented Reality | 23 | 40 |
| 190 | Q1 | ACM Symposium on Access Control Models and Technologies (SACMAT) | 23 | 38 |
| 191 | Q1 | International IFIP TC 6 Conference on Networking | 23 | 37 |
| 191 | Q1 | International Conference on Automated Deduction (CADE) | 23 | 37 |
| 193 | Q1 | IEEE Custom Integrated Circuits Conference, CICC | 23 | 33 |
| 193 | Q1 | IEEE International Conference on Advanced Video and Signal-Based Surveillance (AVSS) | 23 | 33 |
| 193 | Q1 | IEEE Radio Frequency Integrated Circuits Symposium | 23 | 33 |
| 193 | Q1 | IEEE-RAS International Conference on Humanoid Robots (Humanoids) | 23 | 33 |
| 197 | Q1 | European Conference on Information Systems | 23 | 30 |
| 197 | Q1 | International Conference on Mobile Data Management | 23 | 30 |
| 197 | Q1 | Pacific Symposium on Biocomputing | 23 | 30 |
| 200 | Q1 | IEEE International Conference on Multimedia and Expo | 23 | 28 |
| 201 | Q1 | International Conference on Aspect-Oriented Software Development (AOSD) | 22 | 39 |
| 202 | Q1 | IEEE International Conference on Network Protocols | 22 | 37 |
| 203 | Q1 | ACM/IFIP/USENIX International Conference on Middleware | 22 | 34 |
| 203 | Q1 | Pacific-Asia Conference on Advances in Knowledge Discovery and Data Mining | 22 | 34 |
| 205 | Q1 | IEEE International Conference on Program Comprehension | 22 | 33 |
| 206 | Q1 | IEEE International Conference on Automatic Face & Gesture Recognition and Workshops | 22 | 32 |
| 206 | Q1 | IEEE International Conference on Systems, Man and Cybernetics | 22 | 32 |
| 206 | Q1 | IEEE International Symposium on Industrial Electronics | 22 | 32 |
| 206 | Q1 | International Conference on Research in Computational Molecular Biology | 22 | 32 |
| 210 | Q1 | ACM International Conference on Embedded Software | 22 | 31 |
| 210 | Q1 | ACM Multimedia Systems Conference (MMSys) | 22 | 31 |
| 210 | Q1 | Americas Conference on Information Systems (AMCIS) | 22 | 31 |
| 210 | Q1 | International Conference on Database Theory | 22 | 31 |
| 210 | Q1 | Theory and Applications of Satisfiability Testing | 22 | 31 |
| 215 | Q1 | IEEE Conference on Computational Complexity | 22 | 30 |
| 216 | Q1 | International Conference on Field Programmable Logic and Applications | 22 | 29 |
| 217 | Q1 | IEEE International Symposium on Robot and Human Interactive Communication | 22 | 28 |
| 217 | Q1 | IEEE/WIC/ACM International Conference on Web Intelligence and Intelligent Agent Technology | 22 | 28 |
| 217 | Q1 | Principles of Knowledge Representation and Reasoning (KR) | 22 | 28 |
| 220 | Q1 | International Conference on Consumer Electronics | 21 | 34 |
| 221 | Q1 | International ACM/IEEE Joint Conference on Digital Libraries | 21 | 33 |
| 222 | Q1 | IEEE/ACM/IFIP International Conference on Hardware/Software Codesign and System Aynthesis | 21 | 32 |
| 222 | Q1 | Verification, Model Checking and Abstract Interpretation (VMCAI) | 21 | 32 |
| 224 | Q1 | IEEE Radar Conference | 21 | 31 |
| 225 | Q1 | European Conference on Software Maintenance and Reengineering | 21 | 29 |
| 225 | Q1 | IEEE Intelligent Transportation Systems Conference | 21 | 29 |
| 225 | Q1 | IEEE International Conference on Peer-to-Peer Computing | 21 | 29 |
| 225 | Q1 | International Workshop on Wearable and Implantable Body Sensor Networks | 21 | 29 |
| 225 | Q1 | Intelligent Virtual Agents | 21 | 29 |
| 230 | Q1 | IEEE Annual Computer Software and Applications Conference Workshops (COMPSACW) | 21 | 28 |
| 230 | Q1 | International Conference on Concurrency Theory (CONCUR) | 21 | 28 |
| 230 | Q1 | Foundations of Digital Games | 21 | 28 |
| 233 | Q1 | IEEE International Conference on Fuzzy Systems (FUZZ) | 21 | 26 |
| 234 | Q1 | IEEE Virtual Reality Conference | 21 | 25 |
| 234 | Q1 | International Conference on Information processing in medical imaging | 21 | 25 |
| 236 | Q1 | IEEE International Symposium on Performance Analysis of Systems and Software | 20 | 51 |
| 237 | Q1 | International Workshop on Semantic Evaluation | 20 | 49 |
| 238 | Q1 | IEEE International Symposium on a World of Wireless, Mobile and Multimedia Networks | 20 | 37 |
| 238 | Q1 | IEEE International Symposium on Modeling, Analysis & Simulation of Computer and Telecommunication Systems, MASCOTS | 20 | 37 |
| 240 | Q1 | ACM Conference on Hypertext and Hypermedia | 20 | 33 |
| 241 | Q1 | IEEE International Conference on Rehabilitation Robotics | 20 | 31 |
| 241 | Q1 | Design Science Research in Information Systems and Technologies (DESRIST) | 20 | 31 |
| 241 | Q1 | International Conference on Advances in Social Networks Analysis and Mining | 20 | 31 |
| 244 | Q1 | IEEE Symposium on VLSI Circuits | 20 | 29 |
| 244 | Q1 | International Conference on Selected areas in cryptography | 20 | 29 |
| 244 | Q1 | Picture Coding Symposium (PCS) | 20 | 29 |
| 247 | Q1 | Annual Meeting of the Special Interest Group on Discourse and Dialogue (SIGDIAL) | 20 | 28 |
| 247 | Q1 | IEEE International Geoscience and Remote Sensing Symposium | 20 | 28 |
| 247 | Q1 | IEEE Symposium on Field-Programmable Custom Computing Machines | 20 | 28 |
| 247 | Q1 | International Symposium on Modeling and Optimization in Mobile, Ad Hoc, and Wireless Networks and Workshops | 20 | 28 |
| 251 | Q1 | IEEE Pacific Visualization Symposium | 20 | 27 |
| 251 | Q1 | IEEE Workshop on Automatic Speech Recognition & Understanding | 20 | 27 |
| 251 | Q1 | Workshop on Statistical Machine Translation | 20 | 27 |
| 254 | Q1 | European Conference on Antennas and Propagation | 20 | 26 |
| 254 | Q1 | International Conference on Computer Communications and Networks (ICCCN) | 20 | 26 |
| 254 | Q1 | International Conference on Fundamental Approaches to Software Engineering | 20 | 26 |
| 254 | Q1 | International Symposium on Symbolic and Algebraic Computation | 20 | 26 |
| 258 | Q1 | IEEE International Conference on Distributed Computing in Sensor Systems | 20 | 25 |
| 259 | Q1 | ACM International Conference on Supporting group work | 19 | 36 |
| 260 | Q1 | Conference on Image and Video Retrieval | 19 | 33 |
| 261 | Q1 | International Wireless Communications and Mobile Computing Conference (IWCMC) | 19 | 31 |
| 261 | Q1 | International Workshop on Quality of Multimedia Experience | 19 | 31 |
| 263 | Q1 | International Conference on Foundations of Software Science and Computation Structures (FoSSaCS) | 19 | 30 |
| 264 | Q1 | Artificial Intelligence in Education | 19 | 29 |
| 265 | Q1 | IEEE Military Communications Conference | 19 | 28 |
| 265 | Q1 | International Communication Systems and Networks and Workshops COMSNETS | 19 | 28 |
| 267 | Q1 | Conference on Local Computer Networks (LCN) | 19 | 27 |
| 267 | Q1 | Symposium/Workshop on Experimental and Efficient Algorithms | 19 | 27 |
| 269 | Q1 | European Conference on Evolutionary Computation, Machine Learning and Data Mining in Bioinformatics | 19 | 26 |
| 269 | Q1 | IEEE International Conference on Computer and Information Technology (CIT) | 19 | 26 |
| 269 | Q1 | IEEE Statistical Signal Processing Workshop (SSP) | 19 | 26 |
| 269 | Q1 | International Conference on Web Engineering (ICWE) | 19 | 26 |
| 273 | Q1 | AMIA Symposium | 19 | 25 |
| 273 | Q1 | International Symposium on Algorithms and Computation (ISAAC) | 19 | 25 |
| 273 | Q1 | Static Analysis (WSA/SAS) | 19 | 25 |
| 276 | Q1 | International Electron Devices Meeting | 19 | 24 |
| 277 | Q1 | International ACM Conference on Assistive Technologies (Assets) | 19 | 23 |
| 277 | Q1 | Approximation Algorithms for Combinatorial Optimization (APPROX) | 19 | 23 |
| 277 | Q1 | Mathematical Foundations of Computer Science (MFCS) | 19 | 23 |
| 280 | Q1 | IEEE International Conference on Robotics and Biomimetics | 19 | 22 |
| 281 | Q1 | IEEE International Conference on Smart Grid Communications (SmartGridComm) | 19 | 21 |
| 282 | Q1 | IEEE International Symposium on Workload Characterization | 18 | 39 |
| 283 | Q1 | VLSI Test Symposium (VTS) | 18 | 29 |
| 284 | Q1 | Logic Programming and Non-Monotonic Reasoning (LPNMR) | 18 | 28 |
| 285 | Q1 | IEEE Workshop on Signal Processing Advances in Wireless Communications | 18 | 27 |
| 285 | Q1 | International IEEE Enterprise Distributed Object Computing Conference, EDOC | 18 | 27 |
| 287 | Q1 | European Conference on Artificial Intelligence (ECAI) | 18 | 26 |
| 287 | Q1 | IEEE International Symposium on Power Line Communications and Its Applications | 18 | 26 |
| 287 | Q1 | Workshop on Applications of Computer Vision (WACV) | 18 | 26 |
| 290 | Q1 | IEEE International Advance Computing Conference | 18 | 25 |
| 290 | Q1 | IEEE International Power Electronics and Motion Control Conference | 18 | 25 |
| 290 | Q1 | IEEE Symposium on Computers and Communications (ISCC) | 18 | 25 |
| 290 | Q1 | International Symposium on Software Reliability Engineering | 18 | 25 |
| 290 | Q1 | WICSA/ECSA Joint Working IEEE/IFIP Conference on Software Architecture | 18 | 25 |
| 290 | Q1 | International Conference on Conceptual Modeling | 18 | 25 |
| 296 | Q1 | International Solid-State Sensors, Actuators and Microsystems Conference (TRANSDUCERS) | 18 | 24 |
| 297 | Q1 | IEEE International Conference on Advanced Learning Technologies | 18 | 23 |
| 297 | Q1 | IEEE International Conference on Development and Learning (ICDL) | 18 | 23 |
| 297 | Q1 | IEEE International Conference on Global Software Engineering | 18 | 23 |
| 297 | Q1 | IEEE Workshop on Applications of Signal Processing to Audio and Acoustics | 18 | 23 |
| 297 | Q1 | IFIP/IEEE International Symposium on Integrated Network Management | 18 | 23 |
| 302 | Q2 | International Symposium on Distributed Computing (DISC) | 18 | 22 |
| 303 | Q2 | International Conference on VLSI Design | 18 | 21 |
| 304 | Q2 | IEEE International Conference on Computer Design, ICCD | 18 | 20 |
| 304 | Q2 | International Symposium on Quality Electronic Design | 18 | 20 |
| 306 | Q2 | IEEE International Conference on Computer Science and Information Technology (ICCSIT) | 17 | 32 |
| 307 | Q2 | Symposium on Reliable Distributed Systems (SRDS) | 17 | 29 |
| 308 | Q2 | IEEE/ACM International Conference on Grid Computing | 17 | 28 |
| 309 | Q2 | ACM Conference on Designing Interactive Systems | 17 | 27 |
| 309 | Q2 | Rewriting Techniques and Applications (RTA) | 17 | 27 |
| 311 | Q2 | DAGM Symposium for Pattern Recognition | 17 | 26 |
| 312 | Q2 | ACM International Conference on Modeling, Analysis and Simulation of Wireless and Mobile Systems | 17 | 25 |
| 312 | Q2 | IEEE Symposium on Mass Storage Systems and Technologies | 17 | 25 |
| 312 | Q2 | International Symposium on Physical Design | 17 | 25 |
| 315 | Q2 | European Microwave Conference | 17 | 24 |
| 315 | Q2 | IEEE Conference on Industrial Electronics and Applications | 17 | 24 |
| 315 | Q2 | IEEE International Conference on Industrial Informatics | 17 | 24 |
| 315 | Q2 | IEEE Vehicular Networking Conference (VNC) | 17 | 24 |
| 315 | Q2 | International Conference on Compilers, Architecture, and Synthesis for Embedded Systems (CASES) | 17 | 24 |
| 320 | Q2 | Artificial Intelligence and Interactive Digital Entertainment Conference | 17 | 22 |
| 320 | Q2 | International Conference on Machine Learning and Cybernetics | 17 | 22 |
| 320 | Q2 | International Conference on Wireless Communications, Networking and Mobile Computing | 17 | 22 |
| 320 | Q2 | Workshop on Advanced Issues of E-Commerce and Web/based Information Systems (WECWIS) | 17 | 22 |
| 320 | Q2 | World Congress on Nature & Biologically Inspired Computing, NaBIC | 17 | 22 |
| 325 | Q2 | IEEE Asian Solid-State Circuits Conference, A-SSCC | 17 | 21 |
| 325 | Q2 | IEEE International Symposium on Power Semiconductor Devices and ICs | 17 | 21 |
| 325 | Q2 | International Conference on Database Systems for Advanced Applications | 17 | 21 |
| 328 | Q2 | International Conference on Parallel and Distributed Systems | 17 | 20 |
| 328 | Q2 | Symposium on Graph Drawing (GD) | 17 | 20 |
| 330 | Q2 | IEEE International Conference on Wireless and Mobile Computing Networking and Communications | 17 | 19 |
| 331 | Q2 | Information Security Conference / Workshop (ISC/ISW) | 16 | 36 |
| 332 | Q2 | ACM SIGCHI Symposium on Engineering interactive computing systems | 16 | 29 |
| 333 | Q2 | International Conference on Wireless Communication, Vehicular Technology, Information Theory and Aerospace & Electronic Systems Technology Wireless VITAE | 16 | 28 |
| 333 | Q2 | International Conference on Wireless Communications and Signal Processing | 16 | 28 |
| 333 | Q2 | International Conference on Wireless On-Demand Network Systems and Services | 16 | 28 |
| 336 | Q2 | International Conference on Advanced Robotics | 16 | 27 |
| 336 | Q2 | International Conference on Distributed Computing Systems Workshops, ICDCSW | 16 | 27 |
| 338 | Q2 | Annual Conference on Computer Graphics (SIGGRAPH) | 16 | 26 |
| 338 | Q2 | IEEE/AIAA Digital Avionics Systems Conference (DASC) | 16 | 26 |
| 340 | Q2 | ACM Conference on Data and application security and privacy | 16 | 25 |
| 340 | Q2 | IEEE International Conference on Power Electronics and ECCE Asia (ICPE & ECCE) | 16 | 25 |
| 340 | Q2 | IEEE International Workshop on Information Forensics and Security | 16 | 25 |
| 340 | Q2 | International Conference on Digital Signal Processing | 16 | 25 |
| 344 | Q2 | IEEE Symposium on VLSI Technology | 16 | 24 |
| 345 | Q2 | ACM/IEEE International Conference on Distributed Smart Cameras | 16 | 23 |
| 345 | Q2 | IEEE International Symposium on Wearable Computers | 16 | 23 |
| 345 | Q2 | International Conference on Complex, Intelligent and Software Intensive Systems | 16 | 23 |
| 345 | Q2 | International Conference on Logic Programming | 16 | 23 |
| 349 | Q2 | International Conference on Computational Linguistics and Intelligent Text Processing | 16 | 22 |
| 349 | Q2 | International Conference on Network and System Security | 16 | 22 |
| 349 | Q2 | International Conference on Quality Software | 16 | 22 |
| 349 | Q2 | WRI International Conference on Communications and Mobile Computing | 16 | 22 |
| 349 | Q2 | Automated Technology for Verification and Analysis | 16 | 22 |
| 354 | Q2 | Conference on Web Accessibility | 16 | 21 |
| 354 | Q2 | European Conference on Software Architecture | 16 | 21 |
| 354 | Q2 | IEEE International Conference on Communication Software and Networks (ICCSN) | 16 | 21 |
| 354 | Q2 | IEEE Symposium on Visual Languages and Human-Centric Computing | 16 | 21 |
| 354 | Q2 | IEEE Workshop on Multimedia Signal Processing | 16 | 21 |
| 354 | Q2 | International Conference on Parallel problem solving from nature | 16 | 21 |
| 354 | Q2 | International Conference on Ultra Modern Telecommunications & Workshops ICUMT | 16 | 21 |
| 354 | Q2 | Nordic Conference on Human-Computer Interaction: Extending Boundaries | 16 | 21 |
| 354 | Q2 | Logic Programming and Automated Reasoning (RCLP/LPAR) | 16 | 21 |
| 354 | Q2 | International Workshop on Quality of Service (IWQoS) | 16 | 21 |
| 364 | Q2 | Joint Conference on Innovation and Technology in Computer Science Education | 16 | 20 |
| 364 | Q2 | Linguistic Annotation Workshop | 16 | 20 |
| 366 | Q2 | ACM Great Lakes Symposium on VLSI | 16 | 19 |
| 366 | Q2 | Australasian Conference on Information Security and Privacy (ACISP) | 16 | 19 |
| 366 | Q2 | IEEE International Conference on Software Engineering and Formal Methods | 16 | 19 |
| 366 | Q2 | International Conference on Requirements Engineering: Foundation for Software Quality | 16 | 19 |
| 370 | Q2 | IEEE Symposium on Computational Intelligence and Data Mining | 15 | 30 |
| 371 | Q2 | Data Compression Conference, DCC | 15 | 26 |
| 372 | Q2 | Electronics Packaging Technology Conference | 15 | 25 |
| 372 | Q2 | IEEE Computer Society Annual Symposium on VLSI | 15 | 25 |
| 372 | Q2 | International Conference on Intelligent Tutoring Systems | 15 | 25 |
| 372 | Q2 | International Conference on Independent Component Analysis and Signal Separation | 15 | 25 |
| 376 | Q2 | WSEAS International Conference on Automatic control, modelling & simulation | 15 | 24 |
| 376 | Q2 | Recent Advances in Natural Language Processing - RANLP | 15 | 24 |
| 376 | Q2 | International Conference on Knowledge capture | 15 | 24 |
| 379 | Q2 | IEEE International Conference on Embedded and Real-Time Computing Systems and Applications (RTCSA) | 15 | 23 |
| 379 | Q2 | IEEE International Midwest Symposium on Circuits and Systems | 15 | 23 |
| 379 | Q2 | International Conference on Communications and Networking in China | 15 | 23 |
| 379 | Q2 | International Conference on Intelligent Information Hiding and Multimedia Signal Processing | 15 | 23 |
| 379 | Q2 | International Conference on Internet and Web Applications and Services | 15 | 23 |
| 379 | Q2 | International Symposium on Wireless Communication Systems | 15 | 23 |
| 385 | Q2 | IEEE International Conference on Bioinformatics and Biomedicine Workshops (BIBMW) | 15 | 22 |
| 385 | Q2 | IEEE International Symposium on Consumer Electronics | 15 | 22 |
| 385 | Q2 | International Conference on Advances in spatial and temporal databases | 15 | 22 |
| 385 | Q2 | International Symposium on Power Electronics, Electrical Drives, Automation and Motion | 15 | 22 |
| 385 | Q2 | ICCS [International Conference on Computational Science] | 15 | 22 |
| 390 | Q2 | ACM International Conference on Computing Frontiers | 15 | 21 |
| 390 | Q2 | International Conference on Image Analysis and Recognition | 15 | 21 |
| 390 | Q2 | International Joint Conference on Natural Language Processing | 15 | 21 |
| 390 | Q2 | International Symposium on Visual Computing | 15 | 21 |
| 394 | Q2 | European Solid State Device Research Conference, ESSDERC | 15 | 20 |
| 394 | Q2 | IEEE International Conference on Digital Ecosystems and Technologies, DEST | 15 | 20 |
| 394 | Q2 | IEEE International Conference on Intelligent Computing and Intelligent Systems | 15 | 20 |
| 394 | Q2 | International Conference on Bioinformatics and Biomedical Engineering | 15 | 20 |
| 394 | Q2 | Symposium on Haptic Interfaces for Virtual Environment and Teleoperator Systems | 15 | 20 |
| 399 | Q2 | Conference on Current Trends in Theory and Practice of Computer Science | 15 | 19 |
| 399 | Q2 | IEEE Education Engineering (EDUCON) | 15 | 19 |
| 399 | Q2 | IEEE International Conference on Engineering of Complex Computer Systems | 15 | 19 |
| 399 | Q2 | IEEE International Symposium on Network Computing and Applications | 15 | 19 |
| 399 | Q2 | IEEE/ASME International Conference on Advanced Intelligent Mechatronics | 15 | 19 |
| 399 | Q2 | International Conference on Machine Learning and Applications | 15 | 19 |
| 399 | Q2 | International Symposium on Wireless Pervasive Computing | 15 | 19 |
| 399 | Q2 | International Work-Conference on Artificial and Natural Neural Networks (IWANN) | 15 | 19 |
| 399 | Q2 | International Workshop on Education Technology and Computer Science | 15 | 19 |
| 399 | Q2 | Power Electronics and Motion Control Conference | 15 | 19 |
| 409 | Q2 | European Conference on Genetic Programming | 15 | 18 |
| 409 | Q2 | IEEE International Conference on Semantic Computing | 15 | 18 |
| 409 | Q2 | IEEE Symposium on 3D User Interfaces | 15 | 18 |
| 409 | Q2 | International Conference of Distributed Computing and Networking | 15 | 18 |
| 413 | Q2 | EUROMICRO Conference on Software Engineering and Advanced Applications | 15 | 17 |
| 413 | Q2 | IEEE Spoken Language Technology Workshop (SLT) | 15 | 17 |
| 413 | Q2 | International Conference on Natural Computation | 15 | 17 |
| 416 | Q2 | IEEE International Conference on Telecommunications (ICT) | 15 | 16 |
| 416 | Q2 | International Congress on Image and Signal Processing, CISP | 15 | 16 |
| 418 | Q2 | International Conference on Network-Based Information Systems | 14 | 26 |
| 419 | Q2 | International Conference on Artificial Neural Networks | 14 | 25 |
| 419 | Q2 | International Conference on Scale Space and Variational Methods in Computer Vision | 14 | 25 |
| 421 | Q2 | IEEE Asia-Pacific Services Computing Conference, APSCC | 14 | 22 |
| 422 | Q2 | Chinese Control and Decision Conference | 14 | 21 |
| 422 | Q2 | IEEE International Conference on Mechatronics | 14 | 21 |
| 422 | Q2 | IEEE International Symposium on Multimedia | 14 | 21 |
| 422 | Q2 | International Conference on Automotive User Interfaces and Interactive Vehicular Applications | 14 | 21 |
| 422 | Q2 | International Conference on Networks Security, Wireless Communications and Trusted Computing | 14 | 21 |
| 422 | Q2 | International Symposium on Personal, Indoor and Mobile Radio Communications Workshops (PIMRC Workshops) | 14 | 21 |
| 428 | Q2 | ACM Symposium on Document Engineering | 14 | 20 |
| 428 | Q2 | International Conference on Power Electronics and Drive Systems | 14 | 20 |
| 428 | Q2 | Data Warehousing and Knowledge Discovery (DaWaK) | 14 | 20 |
| 428 | Q2 | Web Information Systems Engineering (WISE) | 14 | 20 |
| 428 | Q2 | Frontiers in Education Conference | 14 | 20 |
| 433 | Q2 | Antennas & Propagation Conference | 14 | 19 |
| 433 | Q2 | Asian Conference on Programming languages and systems | 14 | 19 |
| 433 | Q2 | IAPR International Workshop on Document Analysis Systems | 14 | 19 |
| 433 | Q2 | International Conference on Biomedical Engineering and Informatics | 14 | 19 |
| 433 | Q2 | International Conference on Hybrid Artificial Intelligence Systems | 14 | 19 |
| 433 | Q2 | International Conference on Intelligent Sensors, Sensor Networks and Information Processing | 14 | 19 |
| 433 | Q2 | Symposium on VLSI Circuits (VLSIC) | 14 | 19 |
| 433 | Q2 | International Conference on Emerging Trends in Engineering and Technology | 14 | 19 |
| 441 | Q2 | Australian Computer-Human Interaction Conference | 14 | 18 |
| 441 | Q2 | Conference on Multimedia Modeling | 14 | 18 |
| 441 | Q2 | IEEE/ACS International Conference on Computer Systems and Applications | 14 | 18 |
| 441 | Q2 | International Conference on Artificial Intelligence and Law | 14 | 18 |
| 441 | Q2 | International Conference on Electronics Computer Technology (ICECT) | 14 | 18 |
| 441 | Q2 | International Conference on Enterprise Information Systems (ICEIS) | 14 | 18 |
| 441 | Q2 | International Conference on Fuzzy Systems and Knowledge Discovery | 14 | 18 |
| 441 | Q2 | International Conference on Intelligent System Applications to Power Systems ISAP | 14 | 18 |
| 441 | Q2 | International Conference on Measuring Technology and Mechatronics Automation | 14 | 18 |
| 441 | Q2 | International Conference On Principles Of DIstributed Systems | 14 | 18 |
| 451 | Q2 | ACM International Health Informatics Symposium | 14 | 17 |
| 451 | Q2 | ACM Symposium on Virtual Reality Software and Technology | 14 | 17 |
| 451 | Q2 | Advanced International Conference on Telecommunications | 14 | 17 |
| 451 | Q2 | IEEE International Conference on Tools with Artificial Intelligence | 14 | 17 |
| 451 | Q2 | International Conference on Computer Safety, Reliability, and Security | 14 | 17 |
| 451 | Q2 | International Conference on Electrical Engineering/Electronics, Computer, Telecommunications and Information Technology | 14 | 17 |
| 451 | Q2 | International Conference on Frontiers in Handwriting Recognition | 14 | 17 |
| 451 | Q2 | International Conference on Industrial and Information Systems (IIS) | 14 | 17 |
| 451 | Q2 | International Conference on Information Assurance and Security | 14 | 17 |
| 460 | Q2 | Asia Pacific Microwave Conference | 14 | 16 |
| 460 | Q2 | IEEE International Conference on Intelligence and Security Informatics | 14 | 16 |
| 460 | Q2 | IEEE International Conference on Networking, Sensing and Control | 14 | 16 |
| 460 | Q2 | IEEE International Workshop on Policies for Distributed Systems and Networks | 14 | 16 |
| 460 | Q2 | International Conference on Mechatronics and Automation | 14 | 16 |
| 460 | Q2 | International Symposium on Communications, Control and Signal Processing | 14 | 16 |
| 466 | Q2 | IEEE International Symposium on Computer-Based Medical Systems, CBMS | 14 | 15 |
| 467 | Q2 | IEEE International Working Conference on Source Code Analysis and Manipulation | 13 | 28 |
| 468 | Q2 | International Symposium on Communications and Information Technologies | 13 | 23 |
| 468 | Q2 | Annual IFIP WG 11.3 Conference on Data and applications security and privacy | 13 | 23 |
| 470 | Q2 | Asia-Pacific Software Engineering Conference | 13 | 22 |
| 470 | Q2 | IEEE/ACM International Conference on Utility and Cloud Computing | 13 | 22 |
| 472 | Q2 | IEEE Symposium on Industrial Electronics & Applications | 13 | 21 |
| 472 | Q2 | International ITG Workshop on Smart Antennas (WSA) | 13 | 21 |
| 472 | Q2 | International Software Process Conference / Workshop | 13 | 21 |
| 475 | Q2 | IEEE/ACM International Conference on Formal Methods and Models for Codesign | 13 | 20 |
| 475 | Q2 | International Conference for Internet Technology and Secured Transactions | 13 | 20 |
| 475 | Q2 | International Telecommunications Energy Conference | 13 | 20 |
| 475 | Q2 | Latin American Theoretical INformatics (LATIN) | 13 | 20 |
| 479 | Q2 | European Conference on Mobile Robots | 13 | 19 |
| 479 | Q2 | German Conference on Advances in artificial intelligence | 13 | 19 |
| 479 | Q2 | IEEE Workshop on Signal Processing Systems | 13 | 19 |
| 479 | Q2 | IEEE/RAS-EMBS International Conference on Biomedical Robotics and Biomechatronics | 13 | 19 |
| 479 | Q2 | Algorithmic Learning Theory (ALT) | 13 | 19 |
| 479 | Q2 | International Conference on High Performance Computing | 13 | 19 |
| 485 | Q2 | IEEE International Symposium on Precision Clock Synchronization for Measurement, Control and Communication | 13 | 18 |
| 485 | Q2 | International Symposium on Experimental Robotics | 13 | 18 |
| 485 | Q2 | International Workshop on Database and Expert Systems Applications | 13 | 18 |
| 488 | Q2 | European Microwave Integrated Circuits Conference | 13 | 17 |
| 488 | Q2 | IEEE International Symposium on Broadband Multimedia Systems and Broadcasting | 13 | 17 |
| 488 | Q2 | International Conference on Architecture of Computing Systems | 13 | 17 |
| 488 | Q2 | International Conference on Artificial Intelligence and Soft Computing ICAISC | 13 | 17 |
| 488 | Q2 | International Conference on Communication Systems and Network Technologies (CSNT) | 13 | 17 |
| 488 | Q2 | International Conference on Hybrid Intelligent Systems | 13 | 17 |
| 488 | Q2 | International Conference on Information Visualization | 13 | 17 |
| 488 | Q2 | International Conference on ITS Telecommunications | 13 | 17 |
| 488 | Q2 | International Conference on Mathematical morphology and its applications to image and signal processing | 13 | 17 |
| 488 | Q2 | International Conference on Signal Processing Systems | 13 | 17 |
| 488 | Q2 | International Power Electronics and Motion Control Conference | 13 | 17 |
| 488 | Q2 | International Symposium on Robotics - ISR | 13 | 17 |
| 488 | Q2 | Intelligent Data Analysis | 13 | 17 |
| 501 | Q2 | Brazilian Symposium on Software Engineering | 13 | 16 |
| 501 | Q2 | European Radar Conference | 13 | 16 |
| 501 | Q2 | IEEE EUROCON, International Conference on Computer as a Tool | 13 | 16 |
| 501 | Q2 | IEEE/ACIS International Conference on Computer and Information Science ICIS | 13 | 16 |
| 501 | Q2 | International Conference on Database and expert systems applications | 13 | 16 |
| 501 | Q2 | International Conference on Electronics and Information Engineering (ICEIE) | 13 | 16 |
| 507 | Q2 | Asian Conference on Intelligent Information and Database Systems | 13 | 15 |
| 507 | Q2 | IEEE Biomedical Circuits and Systems Conference (BioCAS) | 13 | 15 |
| 507 | Q2 | International Conference on Advanced Computer Theory and Engineering (ICACTE) | 13 | 15 |
| 507 | Q2 | Wireless Telecommunications Symposium, WTS | 13 | 15 |
| 511 | Q2 | IEEE International Symposium on Object Oriented Real-Time Distributed Computing | 13 | 14 |
| 511 | Q2 | International Conference on Agents and Artificial Intelligence (ICAART) | 13 | 14 |
| 511 | Q2 | Topical Meeting on Silicon Monolithic Integrated Circuits in RF Systems | 13 | 14 |
| 514 | Q2 | Asia Communications and Photonics Conference and Exhibition | 12 | 27 |
| 515 | Q2 | International Symposium on Image and Signal Processing and Analysis, ISPA | 12 | 23 |
| 516 | Q2 | International Conference of Soft Computing and Pattern Recognition | 12 | 22 |
| 517 | Q2 | International Conference on Wireless and Mobile Communications | 12 | 21 |
| 517 | Q2 | International Conference on Discovery science | 12 | 21 |
| 519 | Q2 | IEEE Bipolar/BiCMOS Circuits and Technology Meeting (BCTM) | 12 | 20 |
| 519 | Q2 | IEEE Compound Semiconductor Integrated Circuit Symposium (CSICS) | 12 | 20 |
| 519 | Q2 | IEEE International Workshop on Machine Learning for Signal Processing | 12 | 20 |
| 519 | Q2 | NASA/ESA Conference on Adaptive Hardware and Systems, AHS | 12 | 20 |
| 523 | Q2 | IEEE International Workshop on Safety, Security and Rescue Robotics | 12 | 19 |
| 523 | Q2 | International Conference on Advances in Pattern Recognition | 12 | 19 |
| 523 | Q2 | International Conference on Electrical Engineering and Informatics (ICEEI) | 12 | 19 |
| 523 | Q2 | International Conference on Image Analysis and Processing, ICIAP | 12 | 19 |
| 527 | Q2 | Expert Systems | 12 | 18 |
| 527 | Q2 | IEEE International Symposium on Approximate Dynamic Programming and Reinforcement Learning | 12 | 18 |
| 527 | Q2 | IEEE Intersociety Conference on Thermal and Thermomechanical Phenomena in Electronic Systems (ITherm) | 12 | 18 |
| 527 | Q2 | International Conference on Computer Vision Systems (ICVS) | 12 | 18 |
| 527 | Q2 | International Conference on Future Computer and Communication (ICFCC) | 12 | 18 |
| 527 | Q2 | International Conference-Workshop on Compatibility and Power Electronics (CPE) | 12 | 18 |
| 527 | Q2 | Workshop on the Synthesis and Simulation of Living Systems (ALIFE) | 12 | 18 |
| 534 | Q2 | ACM International Conference on Design of Communication | 12 | 17 |
| 534 | Q2 | Device Research Conference | 12 | 17 |
| 534 | Q2 | IEEE International Conference on Trust, Security and Privacy in Computing and Communications (TrustCom) | 12 | 17 |
| 534 | Q2 | IEEE Semiconductor Thermal Measurement and Management Symposium | 12 | 17 |
| 534 | Q2 | International Conference on Computational Intelligence and Security (CIS) | 12 | 17 |
| 534 | Q2 | International Conference on Optimization of Electrical and Electronic Equipment | 12 | 17 |
| 534 | Q2 | International Conference on Signal Processing | 12 | 17 |
| 534 | Q2 | International Conference on Theory of Information Retrieval: Advances in Information Retrieval Theory | 12 | 17 |
| 534 | Q2 | International Conferences on Advances in Computer-Human Interactions | 12 | 17 |
| 534 | Q2 | International Symposium on Turbo Codes and Iterative Information Processing | 12 | 17 |
| 534 | Q2 | The European Symposium on Artificial Neural Networks | 12 | 17 |
| 534 | Q2 | International Symposium on Temporal Representation and Reasoning | 12 | 17 |
| 546 | Q2 | Conference on Information technology education | 12 | 16 |
| 546 | Q2 | IC3K - International Joint Conference on Knowledge Discovery, Knowledge Engineering and Knowledge Management | 12 | 16 |
| 546 | Q2 | IEEE International Conference on Secure Software Integration and Reliability Improvement SSIRI | 12 | 16 |
| 546 | Q2 | IEEE International Symposium on Power Electronics for Distributed Generation Systems | 12 | 16 |
| 546 | Q2 | International Conference on Computational Intelligence and Software Engineering, CiSE | 12 | 16 |
| 546 | Q2 | International Conference on Computational Intelligence, Communication Systems and Networks | 12 | 16 |
| 546 | Q2 | International Conference on Computer Science and Education (ICCSE) | 12 | 16 |
| 546 | Q2 | International Conference on Evaluation and Assessment in Software Engineering | 12 | 16 |
| 546 | Q2 | International Conference on Information security applications | 12 | 16 |
| 546 | Q2 | International Conference on Information Systems Security | 12 | 16 |
| 546 | Q2 | International Conference on Information Technology and Applications in Biomedicine | 12 | 16 |
| 546 | Q2 | International Conference on Information, Communications & Signal Processing | 12 | 16 |
| 546 | Q2 | International Conference on Mobile Ad-hoc and Sensor Networks | 12 | 16 |
| 546 | Q2 | International Workshop on Cognitive Information Processing | 12 | 16 |
| 546 | Q2 | Workshop on Algorithms and Data Structures (WADS) | 12 | 16 |
| 546 | Q2 | WRI Global Congress on Intelligent Systems, GCIS | 12 | 16 |
| 562 | Q2 | IEEE International Conference and Workshops on the Engineering of Computer-Based Systems | 12 | 15 |
| 562 | Q2 | IEEE International Conference on Granular Computing | 12 | 15 |
| 562 | Q2 | IEEE International Conference on Internet Multimedia Services Architecture and Applications (IMSAA) | 12 | 15 |
| 562 | Q2 | IET International Conference on Power Electronics, Machines and Drives | 12 | 15 |
| 562 | Q2 | International Conference on Multimedia Information Networking and Security | 12 | 15 |
| 562 | Q2 | International Conference on Networking, Architecture, and Storage | 12 | 15 |
| 562 | Q2 | International Conference on Software Engineering Advances | 12 | 15 |
| 562 | Q2 | International Symposium on Autonomous Decentralized Systems | 12 | 15 |
| 562 | Q2 | International Topical Meeting on & Microwave Photonics Conference, Asia-Pacific, MWP/APMP | 12 | 15 |
| 562 | Q2 | International Workshop on Image Analysis for Multimedia Interactive Services | 12 | 15 |
| 562 | Q2 | Workshop on Hyperspectral Image and Signal Processing: Evolution in Remote Sensing (WHISPERS) | 12 | 15 |
| 573 | Q2 | IEEE International Conference on Bioinformatics and Biomedicine | 12 | 14 |
| 573 | Q2 | IEEE International Conference on e-Health Networking Applications and Services | 12 | 14 |
| 573 | Q2 | IEEE International Symposium on Design and Diagnostics of Electronic Circuits & Systems | 12 | 14 |
| 573 | Q2 | IEEE Malaysia International Conference on Communications (MICC) | 12 | 14 |
| 573 | Q2 | International Conference on Artificial Intelligence and Computational Intelligence, AICI | 12 | 14 |
| 573 | Q2 | International Conference on Communications, Circuits and Systems, ICCCAS | 12 | 14 |
| 573 | Q2 | International Conference on Education Technology and Computer | 12 | 14 |
| 573 | Q2 | International Conference on Neural Information Processing | 12 | 14 |
| 573 | Q2 | International Journal of RF and Microwave Computer-Aided Engineering | 12 | 14 |
| 573 | Q2 | Workshop on Graph-Theoretic Concepts in Computer Science (WG) | 12 | 14 |
| 573 | Q2 | International Conference on Computing and Combinatorics | 12 | 14 |
| 573 | Q2 | Text, Speech and Dialogue (TSD) | 12 | 14 |
| 585 | Q2 | IEEE International Conference on Electro/Information Technology | 12 | 13 |
| 585 | Q2 | International Conference on Broadband, Wireless Computing, Communication and Applications | 12 | 13 |
| 585 | Q2 | International Conference on Thermal, Mechanical and Multiphysics Simulation and Experiments in Micro-Electronics and Micro-Systems | 12 | 13 |
| 585 | Q2 | International Multi-Conference on Systems, Signals and Devices (SSD) | 12 | 13 |
| 589 | Q2 | International Workshop on Computing Education Research | 11 | 25 |
| 590 | Q2 | International Workshop on Spoken Language Translation | 11 | 24 |
| 591 | Q2 | IEEE International Conference on Cognitive Informatics (ICCI) | 11 | 23 |
| 592 | Q2 | International Conference on Computer and Communication Engineering | 11 | 21 |
| 593 | Q2 | International Conference on Software and Data Technologies | 11 | 20 |
| 593 | Q2 | International Workshop on Knowledge Discovery and Data Mining | 11 | 20 |
| 595 | Q2 | European Conference on Circuit Theory and Design, ECCTD | 11 | 19 |
| 595 | Q2 | International Conference on Signal Acquisition and Processing | 11 | 19 |
| 595 | Q2 | International Symposium on Communication Systems Networks and Digital Signal Processing | 11 | 19 |
| 595 | Q2 | Workshop on Cyber Security and Information Intelligence Research | 11 | 19 |
| 599 | Q2 | International Conference on Information Integration and Web-based Applications & Services (IIWAS) | 11 | 18 |
| 599 | Q2 | International Conference on Web-age information management | 11 | 18 |
| 599 | Q2 | International Database Engineering & Applications Symposium | 11 | 18 |
| 599 | Q2 | International Symposium on VLSI Design, Automation and Test | 11 | 18 |
| 603 | Q3 | ACM Symposium on Computing for Development | 11 | 17 |
| 603 | Q3 | Australasian joint Conference on Advances in Artificial Intelligence | 11 | 17 |
| 603 | Q3 | IFIP International Conference on Wireless and Optical Communications Networks | 11 | 17 |
| 603 | Q3 | International Conference on Innovative Mobile and Internet Services in Ubiquitous Computing | 11 | 17 |
| 603 | Q3 | International Workshop on Content-Based Multimedia Indexing | 11 | 17 |
| 603 | Q3 | Artificial Intelligence and Symbolic Computation (AISC) | 11 | 17 |
| 609 | Q3 | Iberian Conference on Pattern Recognition and Image Analysis | 11 | 16 |
| 609 | Q3 | IEEE Digital Signal Processing Workshop | 11 | 16 |
| 609 | Q3 | IEEE International Vacuum Electronics Conference (IVEC) | 11 | 16 |
| 609 | Q3 | International Conference and Workshops on Advances in Information Security and Assurance | 11 | 16 |
| 609 | Q3 | International Conference on Autonomous Robots and Agents | 11 | 16 |
| 609 | Q3 | Joint IFIP Wireless and Mobile Networking Conference (WMNC) | 11 | 16 |
| 615 | Q3 | Brazilian Power Electronics Conference (COBEP) | 11 | 15 |
| 615 | Q3 | Canadian Conference on Computer and Robot Vision (CRV) | 11 | 15 |
| 615 | Q3 | Conference on Control and Fault-Tolerant Systems (SysTol) | 11 | 15 |
| 615 | Q3 | IEEE Conference on Technologies for Homeland Security | 11 | 15 |
| 615 | Q3 | IEEE International Conference on Electronics, Circuits and Systems | 11 | 15 |
| 615 | Q3 | IEEE-CS Conference on Software Engineering Education and Training (CSEE&T) | 11 | 15 |
| 615 | Q3 | International Conference on Electronic Measurement & Instruments | 11 | 15 |
| 615 | Q3 | International Conference on Intelligent Networking and Collaborative Systems (INCoS) | 11 | 15 |
| 615 | Q3 | International Conference on Microelectronics (MIEL) | 11 | 15 |
| 615 | Q3 | International Conference on Web-Based Learning | 11 | 15 |
| 615 | Q3 | International Symposium on Electronic Commerce and Security | 11 | 15 |
| 615 | Q3 | International Symposium on Neural Networks: Advances in Neural Networks | 11 | 15 |
| 615 | Q3 | Wireless and Optical Communications Conference (WOCC) | 11 | 15 |
| 615 | Q3 | Applications of Natural Language to Data Bases (NLDB) | 11 | 15 |
| 629 | Q3 | Canadian Conference on Advances in Artificial Intelligence | 11 | 14 |
| 629 | Q3 | IEEE Conference on Electrical Performance of Electronic Packaging and Systems (EPEPS) | 11 | 14 |
| 629 | Q3 | IEEE International Symposium on Signal Processing and Information Technology | 11 | 14 |
| 629 | Q3 | Industrial Conference on Data Mining | 11 | 14 |
| 629 | Q3 | International Conference on Control, Automation, Robotics and Vision | 11 | 14 |
| 629 | Q3 | International Conference on E-Business and Information System Security, EBISS | 11 | 14 |
| 629 | Q3 | International Conference on Information and Communications Security | 11 | 14 |
| 629 | Q3 | International Conference on Systems, Signals and Image Processing | 11 | 14 |
| 629 | Q3 | International Conference on Web Information Systems and Mining | 11 | 14 |
| 629 | Q3 | Meeting of the North American Fuzzy Information Processing Society, NAFIPS | 11 | 14 |
| 629 | Q3 | Scandinavian Conference on Image Analysis | 11 | 14 |
| 640 | Q3 | Asia-Pacific Conference on Information Processing | 11 | 13 |
| 640 | Q3 | IEEE International Conference on Communication Technology (ICCT) | 11 | 13 |
| 640 | Q3 | IFIP International Conference on Network and Parallel Computing | 11 | 13 |
| 640 | Q3 | International Conference on Computational Intelligence, Modelling and Simulation (CIMSiM) | 11 | 13 |
| 640 | Q3 | International Conference on Computing, Networking and Communications | 11 | 13 |
| 640 | Q3 | International Conference on Intelligent Human-Machine Systems and Cybernetics | 11 | 13 |
| 640 | Q3 | International Symposium on Computer Architecture and High Performance Computing, SBAC-PAD | 11 | 13 |
| 640 | Q3 | Symposia and Workshops on Ubiquitous, Autonomic and Trusted Computing | 11 | 13 |
| 640 | Q3 | The European Conference on Antennas and Propagation: EuCAP | 11 | 13 |
| 640 | Q3 | Advances in Databases and Information Systems (ADBIS) | 11 | 13 |
| 650 | Q3 | IEEE Conference on Robotics, Automation and Mechatronics | 11 | 12 |
| 650 | Q3 | India Software Engineering Conference | 11 | 12 |
| 650 | Q3 | International Conference on Multimedia Computing and Systems | 11 | 12 |
| 650 | Q3 | International Conference on Parallel and Distributed Computing, Applications and Technologies | 11 | 12 |
| 650 | Q3 | Mexican International Conference on Artificial Intelligence (MICAI) | 11 | 12 |
| 655 | Q3 | International Conference on Computer and Communication Technology (ICCCT) | 11 | 11 |
| 655 | Q3 | International Conference on Software, Telecommunications and Computer Networks | 11 | 11 |
| 657 | Q3 | IEEE International Conference on Integrated Circuit Design and Technology and Tutorial, ICICDT | 10 | 21 |
| 658 | Q3 | IEEE International Workshop on Computer Aided Modeling and Design of Communication Links and Networks (CAMAD) | 10 | 20 |
| 658 | Q3 | International Symposium on Pervasive Systems, Algorithms, and Networks | 10 | 20 |
| 660 | Q3 | ACM International Conference on Bioinformatics and Computational Biology | 10 | 18 |
| 660 | Q3 | International Conference on Image Analysis and Signal Processing | 10 | 18 |
| 662 | Q3 | International Conference on Artificial Intelligence (IC-AI) | 10 | 17 |
| 662 | Q3 | International Conference on Hardware and Software: verification and testing | 10 | 17 |
| 662 | Q3 | Proceedings of the Machine Learning and Data Mining in Pattern Recognition (MLDM) | 10 | 17 |
| 662 | Q3 | Workshop on Algorithms and Computation | 10 | 17 |
| 666 | Q3 | Annual ACM Web Science Conference | 10 | 16 |
| 666 | Q3 | IEEE Convention of Electrical and Electronics Engineers in Israel (IEEEI) | 10 | 16 |
| 666 | Q3 | IEEE International Conference on Software Engineering and Service Sciences | 10 | 16 |
| 666 | Q3 | IEEE Singapore International Conference on Communication Systems | 10 | 16 |
| 666 | Q3 | International Semiconductor Device Research Symposium | 10 | 16 |
| 666 | Q3 | International Symposium on Resilient Control Systems | 10 | 16 |
| 666 | Q3 | International Symposium on Symbolic and Numeric Algorithms for Scientific Computing | 10 | 16 |
| 666 | Q3 | International Workshop on Inductive Logic Programming (ILP) | 10 | 16 |
| 674 | Q3 | ACM International Conference on Underwater Networks and Systems | 10 | 15 |
| 674 | Q3 | IEEE International Performance, Computing, and Communications Conference, IPCCC | 10 | 15 |
| 674 | Q3 | International Conference on Legal Knowledge and Information Systems - JURIX | 10 | 15 |
| 674 | Q3 | International Conference on Power Electronics Systems and Applications | 10 | 15 |
| 674 | Q3 | RAIRO-Theoretical Informatics and Applications | 10 | 15 |
| 674 | Q3 | Workshop on Logic, Language, Information and Computation (WoLLIC) | 10 | 15 |
| 680 | Q3 | Asia-Pacific Web Conference | 10 | 14 |
| 680 | Q3 | Edutainment - International Conference on E-learning and Games | 10 | 14 |
| 680 | Q3 | IEEE International Conference on Intelligent Data Acquisition and Advanced Computing Systems (IDAACS) | 10 | 14 |
| 680 | Q3 | IEEE International Conference on Wireless Communications, Networking and Information Security | 10 | 14 |
| 680 | Q3 | IEEE International Workshop on Haptic Audio visual Environments and Games | 10 | 14 |
| 680 | Q3 | IEEE Workshop on Control and Modeling for Power Electronics (COMPEL) | 10 | 14 |
| 680 | Q3 | IEEE/IFIP International Conference on VLSI and System-on-Chip (VLSI-SoC) | 10 | 14 |
| 680 | Q3 | IEEE/SEMI Advanced Semiconductor Manufacturing Conference, ASMC | 10 | 14 |
| 680 | Q3 | International Conference on Cyber-Enabled Distributed Computing and Knowledge Discovery (CyberC) | 10 | 14 |
| 680 | Q3 | International Conference on Electronic Computer Technology | 10 | 14 |
| 680 | Q3 | International Conference on Electronic Packaging Technology & High Density Packaging | 10 | 14 |
| 680 | Q3 | International Conference on Green Circuits and Systems | 10 | 14 |
| 680 | Q3 | International Conference on Systems and Networks Communications | 10 | 14 |
| 680 | Q3 | Joint International Conference on Power Electronics, Drives and Energy Systems (PEDES) | 10 | 14 |
| 680 | Q3 | OptoElectronics and Communications Conference OECC | 10 | 14 |
| 680 | Q3 | Pacific Rim Conference on Multimedia | 10 | 14 |
| 696 | Q3 | ACIS International Conference on Software Engineering, Artificial Intelligence, Networking, and Parallel/Distributed Computing | 10 | 13 |
| 696 | Q3 | Asia Information Retrieval Symposium | 10 | 13 |
| 696 | Q3 | ASME/IFToMM International Conference on Reconfigurable Mechanisms and Robots | 10 | 13 |
| 696 | Q3 | Iberoamerican Congress on Pattern Recognition CIARP | 10 | 13 |
| 696 | Q3 | IEEE Annual Wireless and Microwave Technology Conference | 10 | 13 |
| 696 | Q3 | IEEE Circuits and Systems International Conference on Testing and Diagnosis | 10 | 13 |
| 696 | Q3 | IEEE International Conference on Bioinformatics and Bioengineering, BIBE | 10 | 13 |
| 696 | Q3 | IEEE International Conference on Broadband Network and Multimedia Technology (IC-BNMT) | 10 | 13 |
| 696 | Q3 | IEEE International Conference on Cloud Computing and Intelligence Systems | 10 | 13 |
| 696 | Q3 | IEEE International Conference on Multisensor Fusion and Integration for Intelligent Systems | 10 | 13 |
| 696 | Q3 | IEEE International Conference on Network Infrastructure and Digital Content | 10 | 13 |
| 696 | Q3 | IEEE International Symposium on Computer-Aided Control System Design (CACSD) | 10 | 13 |
| 696 | Q3 | IEEE Pacific Rim Conference on Communications, Computers and Signal Processing PacRim | 10 | 13 |
| 696 | Q3 | International Conference on Adaptive and natural computing algorithms | 10 | 13 |
| 696 | Q3 | International Conference on Computer and Network Technology | 10 | 13 |
| 696 | Q3 | International Conference on Electrical and Control Engineering (ICECE) | 10 | 13 |
| 696 | Q3 | International Conference on Genetic and Evolutionary Computing (ICGEC) | 10 | 13 |
| 696 | Q3 | International Conference on Informatics in Control, Automation and Robotics | 10 | 13 |
| 696 | Q3 | International Conference on the Quality of Information and Communications Technology | 10 | 13 |
| 696 | Q3 | International Journal of Microwave and Wireless Technologies | 10 | 13 |
| 696 | Q3 | International Workshop on Computational Electronics | 10 | 13 |
| 696 | Q3 | World Congress on Software Engineering, WCSE | 10 | 13 |
| 696 | Q3 | Simulation of Adaptive Behavior | 10 | 13 |
| 719 | Q3 | IEEE International Conference on Intelligent Computer Communication and Processing | 10 | 12 |
| 719 | Q3 | IEEE International Conference on Wireless, Mobile and Ubiquitous Technologies in Education | 10 | 12 |
| 719 | Q3 | IEEE International Symposium on Defect and Fault Tolerance of VLSI Systems, DFTVS | 10 | 12 |
| 719 | Q3 | International Asia Conference on Informatics in Control, Automation and Robotics | 10 | 12 |
| 719 | Q3 | International Conference on Anti-counterfeiting, Security, and Identification in Communication | 10 | 12 |
| 719 | Q3 | International Conference on Applied Electronics (AE) | 10 | 12 |
| 719 | Q3 | International Conference on Intelligent Control and Information Processing (ICICIP) | 10 | 12 |
| 719 | Q3 | International Conference on Intelligent Engineering Systems | 10 | 12 |
| 719 | Q3 | International Conference on Multimedia and Ubiquitous Engineering | 10 | 12 |
| 719 | Q3 | International Radar Symposium | 10 | 12 |
| 719 | Q3 | International Symposium on Applied Machine Intelligence and Informatics | 10 | 12 |
| 719 | Q3 | International Symposium on Parallel and Distributed Computing | 10 | 12 |
| 719 | Q3 | International Symposium on Signals, Circuits and Systems | 10 | 12 |
| 719 | Q3 | International Work-Conference on the Interplay Between Natural and Artificial Computation | 10 | 12 |
| 719 | Q3 | ITS World Congress and Exhibition on Intelligent Transport Systems and Services | 10 | 12 |
| 734 | Q3 | Federated Conference on Computer Science and Information Systems | 10 | 11 |
| 734 | Q3 | IEEE MTT-S International Microwave Workshop Series on Innovative Wireless Power Transmission: Technologies, Systems, and Applications (IMWS) | 10 | 11 |
| 734 | Q3 | IEEE Youth Conference on Information, Computing and Telecommunication YC-ICT | 10 | 11 |
| 734 | Q3 | International Conference Image and Vision Computing New Zealand | 10 | 11 |
| 734 | Q3 | International Conference on Advanced Data Mining and Applications | 10 | 11 |
| 734 | Q3 | International Conference on Circuit and Signal Processing | 10 | 11 |
| 734 | Q3 | International Conference on Computational Intelligence and Communication Networks | 10 | 11 |
| 734 | Q3 | International Conference on Computational Intelligence and Natural Computing, CINC | 10 | 11 |
| 734 | Q3 | International Symposium on Intelligent Systems and Informatics | 10 | 11 |
| 734 | Q3 | Pacific Asia Conference on Language, Information, and Computation (PACLIC) | 10 | 11 |
| 734 | Q3 | Symposium on Integrated Circuits and Systems Design | 10 | 11 |
| 745 | Q3 | Biennial Symposium on Communications (QBSC) | 9 | 17 |
| 745 | Q3 | IEEE International Conference on Bio-Inspired Computing: Theories and Applications | 9 | 17 |
| 745 | Q3 | IFIP TC 6/TC International Conference on Communications and Multimedia Security | 9 | 17 |
| 748 | Q3 | Canadian Conference on Computer Science & Software Engineering | 9 | 15 |
| 748 | Q3 | IEEE International Conference on Microwaves, Communications, Antennas and Electronics Systems | 9 | 15 |
| 748 | Q3 | IEEE International Conference on Wireless Information Technology and Systems | 9 | 15 |
| 748 | Q3 | International Conference on Electrical and Electronics Engineering | 9 | 15 |
| 748 | Q3 | International Symposium on Computer Network and Multimedia Technology | 9 | 15 |
| 748 | Q3 | International Symposium on Information Engineering and Electronic Commerce | 9 | 15 |
| 748 | Q3 | Microwave Journal | 9 | 15 |
| 755 | Q3 | European Microelectronics and Packaging Conference | 9 | 14 |
| 755 | Q3 | European Signal Processing Conference (EUSIPCO) | 9 | 14 |
| 755 | Q3 | International Conference on Computer Graphics, Imaging and Visualization, CGIV | 9 | 14 |
| 755 | Q3 | International Conference on Consumer Electronics, Communications and Networks | 9 | 14 |
| 755 | Q3 | International Conference on Image Processing Theory Tools and Applications (IPTA) | 9 | 14 |
| 755 | Q3 | International Conference on Internet Technology and Applications | 9 | 14 |
| 755 | Q3 | International Conference on Signals, Circuits and Systems | 9 | 14 |
| 755 | Q3 | International ICST Mobile Multimedia Communications Conference | 9 | 14 |
| 755 | Q3 | International Workshop on Cellular Neural Networks and Their Applications | 9 | 14 |
| 764 | Q3 | ACM workshop on Performance monitoring and measurement of heterogeneous wireless and wired networks | 9 | 13 |
| 764 | Q3 | IEEE International Conference on Computational Intelligence and Computing Research (ICCIC) | 9 | 13 |
| 764 | Q3 | IEEE International Conference on Signal and Image Processing Applications | 9 | 13 |
| 764 | Q3 | IEEE/IFIP International Conference on Embedded and Ubiquitous Computing | 9 | 13 |
| 764 | Q3 | International Conference on Computer Aided Systems Theory | 9 | 13 |
| 764 | Q3 | International Conference on Industrial Mechatronics and Automation (ICIMA) | 9 | 13 |
| 764 | Q3 | International Joint Conference on Bioinformatics, Systems Biology and Intelligent Computing | 9 | 13 |
| 764 | Q3 | International Symposium on Advanced Networks and Telecommunication Systems | 9 | 13 |
| 764 | Q3 | International Symposium on Knowledge Acquisition and Modeling | 9 | 13 |
| 764 | Q3 | MobiQuitous - International Conference on Mobile and Ubiquitous Systems: Networking and Services | 9 | 13 |
| 764 | Q3 | Portuguese Conference on Artificial Intelligence (EPIA) | 9 | 13 |
| 775 | Q3 | ACM International Workshop on Mobility Management and Wireless Access, MOBIWAC | 9 | 12 |
| 775 | Q3 | IEEE International Conference on Technologies for Practical Robot Applications | 9 | 12 |
| 775 | Q3 | IEEE International Conference on Vehicular Electronics and Safety | 9 | 12 |
| 775 | Q3 | IEEE International Symposium on Diagnostics for Electric Machines, Power Electronics & Drives (SDEMPED) | 9 | 12 |
| 775 | Q3 | IEEE International Symposium on Intelligent Signal Processing | 9 | 12 |
| 775 | Q3 | IEEE International Symposium on IT in Medicine & Education | 9 | 12 |
| 775 | Q3 | International Conference on Computer Supported Education - CSEDU | 9 | 12 |
| 775 | Q3 | International Conference on Computer Vision and Graphics | 9 | 12 |
| 775 | Q3 | International Conference on Devices and Communications (ICDeCom) | 9 | 12 |
| 775 | Q3 | International Conference on Information and Multimedia Technology | 9 | 12 |
| 775 | Q3 | International Conference on Natural Language Processing and Knowledge Engineering | 9 | 12 |
| 775 | Q3 | International Conference on Signal Processing and Communication Systems | 9 | 12 |
| 775 | Q3 | International Conference on Software Technology and Engineering | 9 | 12 |
| 775 | Q3 | International Symposium on VLSI Technology, Systems, and Applications | 9 | 12 |
| 775 | Q3 | WSEAS International Conference on Artificial intelligence, knowledge engineering and data bases | 9 | 12 |
| 775 | Q3 | International Conference on Flexible Query Answering Systems (FQAS) | 9 | 12 |
| 791 | Q3 | ACIS International Conference on Software Engineering Research, Management & Applications | 9 | 11 |
| 791 | Q3 | Asia Communications and Photonics Conference (ACP) | 9 | 11 |
| 791 | Q3 | Electronics System-Integration Technology Conference | 9 | 11 |
| 791 | Q3 | IEEE International Conference on Computational Intelligence for Measurement Systems and Applications, CIMSA | 9 | 11 |
| 791 | Q3 | IEEE International Conference on Computing and Communication Technologies, Research, Innovation, and Vision for the Future (RIVF) | 9 | 11 |
| 791 | Q3 | IEEE International Symposium on Electronic Design, Test and Applications | 9 | 11 |
| 791 | Q3 | IEEE Latin-American Conference on Communications | 9 | 11 |
| 791 | Q3 | International Conference on Communication, Computing & Security | 9 | 11 |
| 791 | Q3 | International Conference on Computer Systems and Technologies | 9 | 11 |
| 791 | Q3 | International Conference on Control, Automation and Systems, ICCAS | 9 | 11 |
| 791 | Q3 | International Conference on the Applications of Digital Information and Web Technologies | 9 | 11 |
| 791 | Q3 | International ICST Conference on Cognitive Radio Oriented Wireless Networks and Communications | 9 | 11 |
| 791 | Q3 | International Symposium on Distributed Computing and Applications to Business, Engineering and Science | 9 | 11 |
| 791 | Q3 | International Symposium on Intelligent Information Technology and Security Informatics | 9 | 11 |
| 791 | Q3 | International Symposium on Intelligent Signal Processing and Communication Systems | 9 | 11 |
| 791 | Q3 | International Symposium on the Physical and Failure Analysis of Integrated Circuits | 9 | 11 |
| 791 | Q3 | International Symposium on Wireless Personal Multimedia Communications | 9 | 11 |
| 791 | Q3 | Power Electronics, Drive Systems and Technologies Conference (PEDSTC) | 9 | 11 |
| 791 | Q3 | The Journal of microwave power and electromagnetic energy: a publication of the International Microwave Power Institute | 9 | 11 |
| 810 | Q3 | IEEE International New Circuits and Systems Conference | 9 | 10 |
| 810 | Q3 | IEEE International Symposium on Computational Intelligence and Informatics (CINTI) | 9 | 10 |
| 810 | Q3 | IEEE Symposium on Computational Intelligence in Bioinformatics and Computational Biology | 9 | 10 |
| 810 | Q3 | International Conference on Intelligent Robotics and Applications | 9 | 10 |
| 814 | Q3 | International Conference on Computers and Devices for Communication | 9 | 9 |
| 815 | Q3 | ACM Symposium on Performance evaluation of wireless ad hoc, sensor, and ubiquitous networks | 8 | 18 |
| 816 | Q3 | International Microsystems Packaging Assembly and Circuits Technology Conference | 8 | 17 |
| 817 | Q3 | IEEE International Carnahan Conference on Security Technology, ICCST | 8 | 16 |
| 817 | Q3 | IEEE Symposium on Computational Intelligence for Security and Defense Applications | 8 | 16 |
| 817 | Q3 | International Conference on Swarm, Evolutionary, and Memetic Computing | 8 | 16 |
| 820 | Q3 | IEEE Topical Meeting on Silicon Monolithic Integrated Circuits in RF Systems (SiRF) | 8 | 15 |
| 820 | Q3 | International Conference on Telecommunications in Modern Satellite, Cable and Broadcasting Services | 8 | 15 |
| 822 | Q3 | Chinese Conference on Pattern Recognition | 8 | 14 |
| 822 | Q3 | Iberian Conference on Information Systems and Technologies | 8 | 14 |
| 822 | Q3 | International Workshop on Fuzzy Logic and Applications | 8 | 14 |
| 825 | Q3 | European Workshop on Visual Information Processing | 8 | 13 |
| 825 | Q3 | International Conference on Computer Communication and Informatics | 8 | 13 |
| 825 | Q3 | International Conference on Virtual Systems and Multimedia | 8 | 13 |
| 825 | Q3 | International Symposium on Electronic System Design | 8 | 13 |
| 829 | Q3 | European Conference on Software Process Improvement | 8 | 12 |
| 829 | Q3 | German Microwave Conference | 8 | 12 |
| 829 | Q3 | Hellenic Conference on Artificial Intelligence (SETN) | 8 | 12 |
| 829 | Q3 | IEEE International Conference on Grey Systems and Intelligent Services | 8 | 12 |
| 829 | Q3 | IEEE International Conference on Microelectronic Test Structures | 8 | 12 |
| 829 | Q3 | International Conference on Advances in Power System Control, Operation and Management | 8 | 12 |
| 829 | Q3 | International Conference on Cloud and Service Computing | 8 | 12 |
| 829 | Q3 | International Conference on Signal Processing and Communications | 8 | 12 |
| 829 | Q3 | International Conference on Solid-State and Integrated Circuit Technology | 8 | 12 |
| 829 | Q3 | International Conference on Theory and Practice of Electronic Governance | 8 | 12 |
| 829 | Q3 | International Wireless Internet Conference | 8 | 12 |
| 829 | Q3 | Joint Conferences on Pervasive Computing (JCPC) | 8 | 12 |
| 841 | Q3 | Biennial Baltic Electronics Conference (BEC) | 8 | 11 |
| 841 | Q3 | Ibero-American Conference on Advances in artificial intelligence | 8 | 11 |
| 841 | Q3 | IEEE International Conference on Computer Science and Automation Engineering (CSAE) | 8 | 11 |
| 841 | Q3 | IEEE International Conference on Virtual Environments, Human-Computer Interfaces and Measurements Systems | 8 | 11 |
| 841 | Q3 | IEEE National Aerospace and Electronics Conference (NAECON) | 8 | 11 |
| 841 | Q3 | Integrated Communications, Navigation and Surveillance Conference | 8 | 11 |
| 841 | Q3 | International Conference on Advances in Computer-Human Interactions | 8 | 11 |
| 841 | Q3 | International Conference on Autonomic and Autonomous Systems, ICAS | 8 | 11 |
| 841 | Q3 | International Conference on Computer Information Systems and Industrial Management Applications | 8 | 11 |
| 841 | Q3 | International Conference on Data Mining (DMIN) | 8 | 11 |
| 841 | Q3 | International Conference on Power Electronics and Intelligent Transportation System (PEITS) | 8 | 11 |
| 841 | Q3 | International Symposium on Computer Communication Control and Automation | 8 | 11 |
| 841 | Q3 | International Symposium on Intelligent Ubiquitous Computing and Education | 8 | 11 |
| 841 | Q3 | International Symposium on Medical Information and Communication Technology | 8 | 11 |
| 841 | Q3 | International Workshop on Database Technology and Applications (DBTA) | 8 | 11 |
| 841 | Q3 | International Workshop on Robotic and Sensors Environments | 8 | 11 |
| 841 | Q3 | Pacific-Asia Conference on Circuits, Communications and System (PACCS) | 8 | 11 |
| 841 | Q3 | Saudi International Electronics, Communications and Photonics Conference (SIECPC) | 8 | 11 |
| 859 | Q3 | Australian Communications Theory Workshop (AusCTW) | 8 | 10 |
| 859 | Q3 | Canadian Workshop on Information Theory | 8 | 10 |
| 859 | Q3 | European Intelligence and Security Informatics Conference | 8 | 10 |
| 859 | Q3 | IEEE Applied Imagery Pattern Recognition Workshop, AIPR | 8 | 10 |
| 859 | Q3 | IEEE Asia Pacific Conference on Circuits and Systems, APCCAS | 8 | 10 |
| 859 | Q3 | IEEE International Conference on Signal Processing, Communications and Computing | 8 | 10 |
| 859 | Q3 | IEEE International Symposium for Design and Technology in Electronic Packaging (SIITME) | 8 | 10 |
| 859 | Q3 | IEEE International Symposium on Applied Computational Intelligence and Informatics (SACI) | 8 | 10 |
| 859 | Q3 | IEEE Region 8 International Conference on Computational Technologies in Electrical and Electronics Engineering (SIBIRCON) | 8 | 10 |
| 859 | Q3 | Indian Conference on Computer Vision, Graphics & Image Processing, ICVGIP | 8 | 10 |
| 859 | Q3 | International Conference on Applied Robotics for the Power Industry | 8 | 10 |
| 859 | Q3 | International Conference on Computer, Mechatronics, Control and Electronic Engineering | 8 | 10 |
| 859 | Q3 | International Conference on Computing Communication and Networking Technologies (ICCCNT) | 8 | 10 |
| 859 | Q3 | International Conference on Image Information Processing (ICIIP) | 8 | 10 |
| 859 | Q3 | International Conference on Intelligent Networks and Intelligent Systems | 8 | 10 |
| 859 | Q3 | International Conference on Methods and Models in Automation and Robotics | 8 | 10 |
| 859 | Q3 | International Conference on Multimedia Technology | 8 | 10 |
| 859 | Q3 | International Conference on Wavelet Analysis and Pattern Recognition | 8 | 10 |
| 859 | Q3 | International Joint Conference on Computational Intelligence | 8 | 10 |
| 859 | Q3 | International Semiconductor Conference | 8 | 10 |
| 859 | Q3 | International Spring Seminar on Electronics Technology | 8 | 10 |
| 859 | Q3 | International Symposium on Information Theory and Its Applications | 8 | 10 |
| 859 | Q3 | International Symposium on Mechatronics and its Applications | 8 | 10 |
| 859 | Q3 | International Symposium on Microwave, Antenna, Propagation and EMC Technologies for Wireless Communications | 8 | 10 |
| 859 | Q3 | International Workshop on Structural and Syntactic Pattern Recognition (SSPR) | 8 | 10 |
| 859 | Q3 | OptoElectronics and Communications Conference | 8 | 10 |
| 859 | Q3 | Pacific Rim International Conference on Artificial Intelligence (PRICAI) | 8 | 10 |
| 886 | Q3 | Asian-Pacific Conference on Synthetic Aperture Radar | 8 | 9 |
| 886 | Q3 | IEEE International Conference on Automation, Quality and Testing, Robotics, AQTR | 8 | 9 |
| 886 | Q3 | IEEE International Conference on Communications Technology and Applications | 8 | 9 |
| 886 | Q3 | IEEE International Conference on Computational Cybernetics | 8 | 9 |
| 886 | Q3 | IEEE International Conference on Digital Game and Intelligent Toy Enhanced Learning | 8 | 9 |
| 886 | Q3 | International Conference on Computer and Communication Technologies in Agriculture Engineering (CCTAE) | 8 | 9 |
| 886 | Q3 | International Conference on e-Education, e-Business, e-Management, and e-Learning | 8 | 9 |
| 886 | Q3 | International Conference on Electric Information and Control Engineering (ICEICE) | 8 | 9 |
| 886 | Q3 | International Conference on Interoperability for Enterprise Software and Applications China | 8 | 9 |
| 886 | Q3 | International Conference on Microwave and Millimeter Wave Technology | 8 | 9 |
| 886 | Q3 | International Conference on MultiMedia and Information Technology | 8 | 9 |
| 886 | Q3 | Koli Calling International Conference on Computing Education Research | 8 | 9 |
| 886 | Q3 | Spring Conference on Computer Graphics | 8 | 9 |
| 899 | Q3 | International Conference on Simulation of Semiconductor Processes and Devices | 8 | 8 |
| 899 | Q3 | International IEEE Conference on Signal-Image Technologies and Internet-Based System | 8 | 8 |
| 899 | Q3 | International Workshop on Advanced Computational Intelligence (IWACI) | 8 | 8 |
| 902 | Q4 | Asia Communications and Photonics Conference | 7 | 15 |
| 902 | Q4 | International Conference on Computer Science and Electronics Engineering | 7 | 15 |
| 902 | Q4 | International Conference on Image and Signal Processing | 7 | 15 |
| 902 | Q4 | International Conference on Image Processing, Computer Vision, & Pattern Recognition | 7 | 15 |
| 906 | Q4 | International Conference on Electronic and Mechanical Engineering and Information Technology (EMEIT) | 7 | 14 |
| 907 | Q4 | International Conference on Electronic Commerce | 7 | 13 |
| 907 | Q4 | International Visual Informatics Conference | 7 | 13 |
| 909 | Q4 | IEEE International Semiconductor Laser Conference (ISLC) | 7 | 12 |
| 909 | Q4 | International Conference on Brain informatics | 7 | 12 |
| 909 | Q4 | Nordic Conference on Human-Computer Interaction: Making Sense Through Design | 7 | 12 |
| 912 | Q4 | Electronics, Robotics and Automotive Mechanics Conference | 7 | 11 |
| 912 | Q4 | International Conference on Bioinformatics and Biomedical Technology (ICBBT) | 7 | 11 |
| 912 | Q4 | International Conference on Educational and Information Technology (ICEIT) | 7 | 11 |
| 912 | Q4 | International Conference on Ground Penetrating Radar | 7 | 11 |
| 912 | Q4 | International Conference on Information Retrieval & Knowledge Management | 7 | 11 |
| 912 | Q4 | International Conference on Intelligent Computing and Cognitive Informatics | 7 | 11 |
| 912 | Q4 | International Conference on Next Generation Web Services Practices | 7 | 11 |
| 912 | Q4 | International Conference on Pervasive Computing and Applications | 7 | 11 |
| 912 | Q4 | International Conference on Signal Processing, Communication, Computing and Networking Technologies | 7 | 11 |
| 912 | Q4 | International Workshop on Software Measurement (IWSM) | 7 | 11 |
| 912 | Q4 | Computer Graphics Theory and Applications | 7 | 11 |
| 923 | Q4 | Catalan Conference on AI | 7 | 10 |
| 923 | Q4 | Euro American Conference on Telematics and Information Systems | 7 | 10 |
| 923 | Q4 | IEEE International Conference on Service Operations and Logistics, and Informatics | 7 | 10 |
| 923 | Q4 | IEEE Symposium on Computers & Informatics (ISCI) | 7 | 10 |
| 923 | Q4 | IITA International Conference on Control, Automation and Systems Engineering | 7 | 10 |
| 923 | Q4 | International Conference on Advances in artificial intelligence: spanish association for artificial intelligence | 7 | 10 |
| 923 | Q4 | International Conference on Communications and Signal Processing (ICCSP) | 7 | 10 |
| 923 | Q4 | International Conference on Computational Advances in Bio and Medical Sciences | 7 | 10 |
| 923 | Q4 | International Conference on Electronic Devices, Systems and Applications (ICEDSA) | 7 | 10 |
| 923 | Q4 | International Conference on Microwave Radar and Wireless Communications | 7 | 10 |
| 923 | Q4 | International Conference on Network Computing and Information Security | 7 | 10 |
| 923 | Q4 | International Conference on New Frontiers in Artificial Intelligence | 7 | 10 |
| 923 | Q4 | International Conference on Power, Control and Embedded Systems | 7 | 10 |
| 923 | Q4 | International IEEE Conference on Intelligent Systems | 7 | 10 |
| 923 | Q4 | International Symposium on Antenna Technology and Applied Electromagnetics & the American Electromagnetics Conference (ANTEM-AMEREM) | 7 | 10 |
| 923 | Q4 | International Workshop on Satellite and Space Communications | 7 | 10 |
| 923 | Q4 | IPCC IEEE International Professional Communication Conference | 7 | 10 |
| 923 | Q4 | Pacific-Asia Workshop on Computational Intelligence and Industrial Application | 7 | 10 |
| 941 | Q4 | Asia Symposium on Quality Electronic Design | 7 | 9 |
| 941 | Q4 | ESA Workshop on Satellite Navigation Technologies and European Workshop on GNSS Signals and Signal Processing (NAVITEC) | 7 | 9 |
| 941 | Q4 | IEEE International Conference on Progress in Informatics and Computing | 7 | 9 |
| 941 | Q4 | IEEE International RF and Microwave Conference, RFM | 7 | 9 |
| 941 | Q4 | IMOC SBMO/IEEE MTT-S International Microwave and Optoelectronics Conference | 7 | 9 |
| 941 | Q4 | International Conference on Artificial Intelligence, Management Science and Electronic Commerce | 7 | 9 |
| 941 | Q4 | International Conference on Computer Applications and Industrial Electronics (ICCAIE) | 7 | 9 |
| 941 | Q4 | International Conference on Digital Content, Multimedia Technology and its Applications | 7 | 9 |
| 941 | Q4 | International Conference on Electronic Visualisation and the Arts | 7 | 9 |
| 941 | Q4 | International Conference on Machine Vision and Human-Machine Interface | 7 | 9 |
| 941 | Q4 | International Conference on Numerical Simulation of Optoelectronic Devices | 7 | 9 |
| 941 | Q4 | International Conference on Signal and Image Processing | 7 | 9 |
| 941 | Q4 | International Conference on Technologies and Applications of Artificial Intelligence | 7 | 9 |
| 941 | Q4 | International Conference on Wireless Networks and Information Systems | 7 | 9 |
| 941 | Q4 | International Symposium on Electronics and Telecommunications | 7 | 9 |
| 941 | Q4 | International Symposium on Plant Growth Modeling, Simulation, Visualization and Applications | 7 | 9 |
| 957 | Q4 | ACM Research in Applied Computation Symposium | 7 | 8 |
| 957 | Q4 | IEEE Conference on Electron Devices and Solid-State Circuits, EDSSC | 7 | 8 |
| 957 | Q4 | IEEE International Conference on Computing, Control and Industrial Engineering (CCIE) | 7 | 8 |
| 957 | Q4 | IEEE Symposium on Web Society (SWS) | 7 | 8 |
| 957 | Q4 | International Conference on Audio Language and Image Processing (ICALIP) | 7 | 8 |
| 957 | Q4 | International Conference on Electronics, Communications and Control (ICECC) | 7 | 8 |
| 957 | Q4 | International Conference on Emerging Trends in Robotics and Communication Technologies | 7 | 8 |
| 957 | Q4 | International Conference on Mechanic Automation and Control Engineering | 7 | 8 |
| 957 | Q4 | International Conference on Mechanical and Electronics Engineering | 7 | 8 |
| 957 | Q4 | International Conference on Software Engineering and Data Mining | 7 | 8 |
| 957 | Q4 | International Symposium on Computational Intelligence and Design | 7 | 8 |
| 957 | Q4 | International Symposium on Integrated Circuits | 7 | 8 |
| 957 | Q4 | International Symposium on Telecommunications | 7 | 8 |
| 957 | Q4 | WSEAS International Conference on Software engineering, parallel and distributed systems | 7 | 8 |
| 971 | Q4 | Asia Pacific Conference on Postgraduate Research in Microelectronics & Electronics PrimeAsia | 7 | 7 |
| 971 | Q4 | DeLFI - Fachtagung e-Learning der Gesellschaft für Informatik | 7 | 7 |
| 971 | Q4 | IEEE Southwest Symposium on Image Analysis & Interpretation (SSIAI) | 7 | 7 |
| 971 | Q4 | International Conference on eHealth, Telemedicine, and Social Medicine | 7 | 7 |
| 971 | Q4 | International Conference on Test and Measurement | 7 | 7 |
| 976 | Q4 | International Vacuum Electron Sources Conference and Nanocarbon (IVESC) | 6 | 21 |
| 977 | Q4 | Conference on Optoelectronic and Microelectronic Materials and Devices (COMMAD) | 6 | 18 |
| 978 | Q4 | ACIS/JNU International Conference on Computers, Networks, Systems and Industrial Engineering | 6 | 12 |
| 978 | Q4 | IEEE International Conference on Information Theory and Information Security | 6 | 12 |
| 978 | Q4 | International Conference on E-Health Networking, Digital Ecosystems and Technologies (EDT) | 6 | 12 |
| 978 | Q4 | WSEAS International Conference on Signal processing, robotics and automation | 6 | 12 |
| 982 | Q4 | International Conference on Emerging Intelligent Data and Web Technologies | 6 | 11 |
| 982 | Q4 | International Conference on Knowledge Management and Knowledge Technologies | 6 | 11 |
| 982 | Q4 | International Conference on Software Engineering & Computer Systems | 6 | 11 |
| 985 | Q4 | Asia-Pacific Conference on Wearable Computing Systems | 6 | 10 |
| 985 | Q4 | Balkan Conference in Informatics | 6 | 10 |
| 985 | Q4 | IEEE International Conference on Computer-Aided Design and Computer Graphics | 6 | 10 |
| 985 | Q4 | IEEE Workshop on Advanced Robotics and its Social Impacts (ARSO) | 6 | 10 |
| 985 | Q4 | International Conference on Computer Science and Network Technology (ICCSNT) | 6 | 10 |
| 985 | Q4 | International Conference on Computing, Electronics and Electrical Technologies | 6 | 10 |
| 985 | Q4 | International Conference on Education and Management Technology (ICEMT) | 6 | 10 |
| 985 | Q4 | International Conference on Instrumentation, Communications, Information Technology, and Biomedical Engineering | 6 | 10 |
| 985 | Q4 | International Conference on Wireless Networks | 6 | 10 |
| 985 | Q4 | Mexican Conference on Pattern Recognition | 6 | 10 |
| 985 | Q4 | Specialist Meeting on Microwave Radiometry and Remote Sensing of the Environment (MicroRad) | 6 | 10 |
| 985 | Q4 | UK Workshop on Computational Intelligence | 6 | 10 |
| 997 | Q4 | IEEE International Conference on Computer-Aided Industrial Design & Conceptual Design | 6 | 9 |
| 997 | Q4 | Indian International Conference on Artificial Intelligence | 6 | 9 |
| 997 | Q4 | International Conference on Communications and Information Technology | 6 | 9 |
| 997 | Q4 | International Conference on Electrical, Control and Computer Engineering | 6 | 9 |
| 997 | Q4 | International Conference on Networking and Distributed Computing | 6 | 9 |
| 997 | Q4 | International Conference on Pervasive Computing Signal Processing and Applications | 6 | 9 |
| 997 | Q4 | International Conference on Supply Chain Management and Information Systems | 6 | 9 |
| 997 | Q4 | International Workshop on Systems, Signal Processing and their Applications | 6 | 9 |
| 997 | Q4 | Spanish Conference on Electron Devices CDE | 6 | 9 |
| 1006 | Q4 | Brazilian Symposium on Human Factors in Computing Systems | 6 | 8 |
| 1006 | Q4 | DoD High Performance Computing Modernization Program Users Group Conference | 6 | 8 |
| 1006 | Q4 | European Conference on Networks and Optical Communications | 6 | 8 |
| 1006 | Q4 | IEEE International Conference on Control System, Computing and Engineering (ICCSCE) | 6 | 8 |
| 1006 | Q4 | IEEE International Conference on Cyber Technology in Automation, Control, and Intelligent Systems | 6 | 8 |
| 1006 | Q4 | IEEE International Symposium on Logistics and Industrial Informatics (LINDI) | 6 | 8 |
| 1006 | Q4 | IITA International Conference on Geoscience and Remote Sensing | 6 | 8 |
| 1006 | Q4 | International Conference on Advances in Computing, Communications and Informatics | 6 | 8 |
| 1006 | Q4 | International Conference on Asian Language Processing | 6 | 8 |
| 1006 | Q4 | International Conference on Communications, Computing and Control Applications | 6 | 8 |
| 1006 | Q4 | International Conference on Electronic Commerce and Business Intelligence, ECBI | 6 | 8 |
| 1006 | Q4 | International Conference on Emerging Trends in Networks and Computer Communications (ETNCC) | 6 | 8 |
| 1006 | Q4 | International Conference on Internet Multimedia Computing and Service | 6 | 8 |
| 1006 | Q4 | International Conference on Mechatronic Science, Electric Engineering and Computer | 6 | 8 |
| 1006 | Q4 | International Conference on Pattern Recognition Applications and Methods | 6 | 8 |
| 1006 | Q4 | International Conference on Process Automation, Control and Computing | 6 | 8 |
| 1006 | Q4 | International Conference System Theory, Control and Computing | 6 | 8 |
| 1006 | Q4 | International Siberian Conference on Control and Communications (SIBCON) | 6 | 8 |
| 1006 | Q4 | International Symposium on Chinese Spoken Language Processing (ISCSLP) | 6 | 8 |
| 1006 | Q4 | International Symposium on Information Processing (ISIP) | 6 | 8 |
| 1006 | Q4 | International Symposium on Systems and Control in Aerospace and Astronautics | 6 | 8 |
| 1006 | Q4 | International Workshop on Signal Design and Its Applications in Communications | 6 | 8 |
| 1028 | Q4 | Brazilian Symposium on Multimedia and the Web | 6 | 7 |
| 1028 | Q4 | IAPR Asian Conference on Pattern Recognition | 6 | 7 |
| 1028 | Q4 | IEEE CIE International Conference on Radar | 6 | 7 |
| 1028 | Q4 | IEEE International Conference on Healthcare Informatics, Imaging and Systems Biology | 6 | 7 |
| 1028 | Q4 | IEEE International Conference on Space Science and Communication (IconSpace) | 6 | 7 |
| 1028 | Q4 | IEEE International Conference on Technology for Education | 6 | 7 |
| 1028 | Q4 | IEEE Nanotechnology Materials and Devices Conference | 6 | 7 |
| 1028 | Q4 | IEEE-APS Topical Conference on Antennas and Propagation in Wireless Communications | 6 | 7 |
| 1028 | Q4 | IEEE-EMBS International Conference on Biomedical and Health Informatics | 6 | 7 |
| 1028 | Q4 | International Conference on Cloud and Green Computing | 6 | 7 |
| 1028 | Q4 | International Conference on Computer Science and Software Engineering | 6 | 7 |
| 1028 | Q4 | International Conference on Educational and Network Technology | 6 | 7 |
| 1028 | Q4 | International Conference on Electronics and Optoelectronics (ICEOE) | 6 | 7 |
| 1028 | Q4 | International Conference on Information Technology Based Higher Education and Training | 6 | 7 |
| 1028 | Q4 | International Conference on Multimedia and Signal Processing | 6 | 7 |
| 1028 | Q4 | International Conference on Optoelectronics and Image Processing | 6 | 7 |
| 1028 | Q4 | International Conference on Pattern Recognition, Informatics and Mobile Engineering | 6 | 7 |
| 1028 | Q4 | International Conference on Signals and Electronic Systems | 6 | 7 |
| 1028 | Q4 | International Conference on Ubiquitous Robots and Ambient Intelligence | 6 | 7 |
| 1028 | Q4 | International Symposium on Intelligence Information Processing and Trusted Computing (IPTC) | 6 | 7 |
| 1028 | Q4 | International Symposium on Parallel Architectures, Algorithms and Programming | 6 | 7 |
| 1028 | Q4 | International Symposium on Signals Systems and Electronics | 6 | 7 |
| 1028 | Q4 | International Telecommunications Network Strategy and Planning Symposium | 6 | 7 |
| 1028 | Q4 | Journal of Microwaves, Optoelectronics and Electromagnetic Applications | 6 | 7 |
| 1028 | Q4 | Mediterranean Microwave Symposium (MMS) | 6 | 7 |
| 1028 | Q4 | Microwaves & RF | 6 | 7 |
| 1028 | Q4 | Symposium on Neural Network Applications in Electrical Engineering (NEUREL) | 6 | 7 |
| 1055 | Q4 | IEEE Joint International Information Technology and Artificial Intelligence Conference (ITAIC) | 6 | 6 |
| 1055 | Q4 | IEEE Latin American Symposium on Circuits and Systems (LASCAS) | 6 | 6 |
| 1055 | Q4 | IEEE Students' Conference on Electrical, Electronics and Computer Science | 6 | 6 |
| 1055 | Q4 | International Conference on Applied Superconductivity and Electromagnetic Devices, ASEMD | 6 | 6 |
| 1055 | Q4 | International Conference on Modern Problems of Radio Engineering Telecommunications and Computer Science | 6 | 6 |
| 1055 | Q4 | International Conference on Recent Advancements in Electrical, Electronics and Control Engineering | 6 | 6 |
| 1055 | Q4 | International Technology, Education and Development Conference | 6 | 6 |
| 1062 | Q4 | International Kharkov Symposium on Physics and Engineering of Microwaves, Millimeter and Submillimeter Waves and Workshop on Terahertz Technologies | 5 | 9 |
| 1063 | Q4 | CSI International Symposium on Artificial Intelligence and Signal Processing | 5 | 8 |
| 1063 | Q4 | IEEE Conference on Computational Intelligence for Financial Engineering & Economics (CIFEr) | 5 | 8 |
| 1063 | Q4 | International Conference on Intelligent Human Computer Interaction (IHCI) | 5 | 8 |
| 1063 | Q4 | International Symposium on Advanced Control of Industrial Processes | 5 | 8 |
| 1067 | Q4 | Chinese Conference on Biometric Recognition | 5 | 7 |
| 1067 | Q4 | IEEE International Conference on Communications in China | 5 | 7 |
| 1067 | Q4 | IEEE International Conference on Spatial Data Mining and Geographical Knowledge Services | 5 | 7 |
| 1067 | Q4 | IEEE International Conference on Teaching, Assessment and Learning for Engineering | 5 | 7 |
| 1067 | Q4 | International Asia-Pacific Conference on Synthetic Aperture Radar | 5 | 7 |
| 1067 | Q4 | International Conference and Seminar on Micro/Nanotechnologies and Electron Devices | 5 | 7 |
| 1067 | Q4 | International Conference on Communication Systems, Networks and Applications (ICCSNA) | 5 | 7 |
| 1067 | Q4 | International Conference on Computer Science and Information Processing | 5 | 7 |
| 1067 | Q4 | International Conference on Emerging Trends in Electronic and Photonic Devices & Systems | 5 | 7 |
| 1067 | Q4 | International Conference on Fluid Power and Mechatronics (FPM) | 5 | 7 |
| 1067 | Q4 | International Conference on Informatics, Electronics & Vision | 5 | 7 |
| 1067 | Q4 | International Conference on Instrumentation, Measurement, Computer, Communication and Control | 5 | 7 |
| 1067 | Q4 | International Conference on Soft Computing for Problem Solving | 5 | 7 |
| 1080 | Q4 | Asia-Pacific Conference on Information Network and Digital Content Security | 5 | 6 |
| 1080 | Q4 | ICMTCE. International Conference on Microwave Technology and Computational Electromagnetics | 5 | 6 |
| 1080 | Q4 | IEEE Conference on Information & Communication Technologies | 5 | 6 |
| 1080 | Q4 | IEEE Electrical Design of Advanced Packaging & Systems Symposium (EDAPS) | 5 | 6 |
| 1080 | Q4 | IEEE International Conference on Microwave Technology & Computational Electromagnetics | 5 | 6 |
| 1080 | Q4 | IEEE International Conference on Signal Processing, Computing and Control | 5 | 6 |
| 1080 | Q4 | International Conference of Fuzzy Information and Engineering | 5 | 6 |
| 1080 | Q4 | International Conference on Advanced Computing, Networking and Security | 5 | 6 |
| 1080 | Q4 | International Conference on Advanced Mechatronic Systems | 5 | 6 |
| 1080 | Q4 | International Conference on Devices, Circuits and Systems | 5 | 6 |
| 1080 | Q4 | International Conference on Research and Innovation in Information Systems | 5 | 6 |
| 1080 | Q4 | International Conference on Systems and Informatics | 5 | 6 |
| 1080 | Q4 | International Symposium on Advanced Packaging Materials (APM) | 5 | 6 |
| 1080 | Q4 | Web Information Systems and Applications Conference (WISA) | 5 | 6 |
| 1094 | Q4 | IEEE International Conference on Computer Science and Automation Engineering | 5 | 5 |
| 1094 | Q4 | IEEE International Conference on e-Health Networking, Applications and Services (Healthcom) | 5 | 5 |
| 1094 | Q4 | IEEE International Workshop on Genomic Signal Processing and Statistics | 5 | 5 |
| 1094 | Q4 | International Conference on Artificial Intelligence and Education (ICAIE) | 5 | 5 |
| 1094 | Q4 | International Conference on Solid-State and Integrated-Circuit Technology | 5 | 5 |
| 1094 | Q4 | OptoElectronics and Communications Conference and Photonics in Switching | 5 | 5 |
| 1094 | Q4 | Quantum Bio-Informatics III From Quantum Information to Bio-Informatics | 5 | 5 |
| 1094 | Q4 | Symposium on Virtual and Augmented Reality | 5 | 5 |
| 1102 | Q4 | IEEE/CPMT International Electronic Manufacturing Technology Symposium (IEMT) | 4 | 10 |
| 1103 | Q4 | International Conference on Optical Communications and Networks | 4 | 9 |
| 1104 | Q4 | IEEE Jordan Conference on Applied Electrical Engineering and Computing Technologies | 4 | 8 |
| 1105 | Q4 | International Conference on Business Management and Electronic Information (BMEI) | 4 | 7 |
| 1105 | Q4 | International Conference on Control, Automation and Systems Engineering (CASE) | 4 | 7 |
| 1105 | Q4 | International Workshop on Pattern Recognition in Neuroimaging | 4 | 7 |
| 1108 | Q4 | Conference on Image and Vision Computing New Zealand | 4 | 6 |
| 1108 | Q4 | International Conference on Practical Applications of Computational Biology & Bioinformatics | 4 | 6 |
| 1108 | Q4 | International Conference on Robot, Vision and Signal Processing | 4 | 6 |
| 1108 | Q4 | International Conference on Virtual Reality and Visualization | 4 | 6 |
| 1108 | Q4 | Sino-foreign-interchange Conference on Intelligent Science and Intelligent Data Engineering | 4 | 6 |
| 1113 | Q4 | IEEE AESS European Conference on Satellite Telecommunications (ESTEL) | 4 | 5 |
| 1113 | Q4 | IEEE International Conference on Networks | 4 | 5 |
| 1113 | Q4 | IEEE International Conference on Semiconductor Electronics | 4 | 5 |
| 1113 | Q4 | IEEE International Conference on Systems Biology | 4 | 5 |
| 1113 | Q4 | International Aegean Conference on Electrical Machines and Power Electronics | 4 | 5 |
| 1113 | Q4 | International Conference on Communication, Electronics and Automation Engineering | 4 | 5 |
| 1113 | Q4 | International Conference on Education and e-Learning Innovations | 4 | 5 |
| 1113 | Q4 | International Conference on Logistics, Informatics and Service Science | 4 | 5 |
| 1113 | Q4 | International Conference on Ubiquitous Computing and Ambient Intelligence | 4 | 5 |
| 1113 | Q4 | International Symposium on Electrical and Electronics Engineering | 4 | 5 |
| 1113 | Q4 | Korea-Japan Joint Workshop on Frontiers of Computer Vision | 4 | 5 |
| 1113 | Q4 | MATEC Web of Conferences | 4 | 5 |
| 1113 | Q4 | National Conference on Electrical, Electronics and Computer Engineering | 4 | 5 |
| 1113 | Q4 | Symposium on Piezoelectricity, Acoustic Waves, and Device Applications (SPAWDA) | 4 | 5 |
| 1127 | Q4 | Computer Science and Electronic Engineering Conference (CEEC) | 4 | 4 |
| 1127 | Q4 | IEEE Global Conference on Consumer Electronics (GCCE) | 4 | 4 |
| 1127 | Q4 | IEEE International Symposium on Next-Generation Electronics | 4 | 4 |
| 1127 | Q4 | IEEE Symposium on Electrical & Electronics Engineering | 4 | 4 |
| 1127 | Q4 | International Conference on Advances in Communication, Network, and Computing | 4 | 4 |
| 1127 | Q4 | International Conference on Communications, Signal Processing, and their Applications | 4 | 4 |
| 1127 | Q4 | International Conference on Industrial Control and Electronics Engineering | 4 | 4 |
| 1127 | Q4 | International Conference on Remote Sensing, Environment and Transportation Engineering | 4 | 4 |
| 1127 | Q4 | International Conference on Technological Advances in Electrical, Electronics and Computer Engineering | 4 | 4 |
| 1127 | Q4 | International Electronic Conference on Synthetic Organic Chemistry | 4 | 4 |
| 1127 | Q4 | Studia i Materialy Polskiego Stowarzyszenia Zarzadzania Wiedza/Studies & Proceedings Polish Association for Knowledge Management | 4 | 4 |
| 1138 | Q4 | Anais do Congresso Nacional Universidade, EAD e Software Livre | 3 | 5 |
| 1138 | Q4 | IIAI International Conference on Advanced Applied Informatics | 3 | 5 |
| 1138 | Q4 | International Conference on Control Engineering and Communication Technology | 3 | 5 |
| 1138 | Q4 | International Symposium on Semiconductor Manufacturing | 3 | 5 |
| 1142 | Q4 | Conference on Software Engineering Education | 3 | 4 |
| 1142 | Q4 | IEEE Asia-Pacific Conference on Antennas and Propagation | 3 | 4 |
| 1142 | Q4 | IEEE International Meeting for Future of Electron Devices, Kansai (IMFEDK) | 3 | 4 |
| 1142 | Q4 | IEEE Symposium on Robotics and Applications | 3 | 4 |
| 1142 | Q4 | International Conference in Electrics, Communication and Automatic Control | 3 | 4 |
| 1142 | Q4 | International Conference on Advanced Semiconductor Devices & Microsystems (ASDAM) | 3 | 4 |
| 1142 | Q4 | International Conference on Antenna Theory and Techniques | 3 | 4 |
| 1142 | Q4 | International Conference on Applied Informatics and Communication | 3 | 4 |
| 1142 | Q4 | International Conference on Computing, Measurement, Control and Sensor Network | 3 | 4 |
| 1142 | Q4 | International Conference on Engineering and Technology Education | 3 | 4 |
| 1142 | Q4 | International Conference on Sciences of Electronics, Technologies of Information and Telecommunications | 3 | 4 |
| 1142 | Q4 | International Symposium on Instrumentation and Measurement, Sensor Network and Automation | 3 | 4 |
| 1142 | Q4 | National Conference on Computer Vision, Pattern Recognition, Image Processing and Graphics | 3 | 4 |
| 1142 | Q4 | SHS Web of Conferences | 3 | 4 |
| 1156 | Q4 | 2012 IEEE Fifth International Conference on Advanced Computational Intelligence (ICACI) | 3 | 3 |
| 1156 | Q4 | Asia-Pacific Conference on Information Theory | 3 | 3 |
| 1156 | Q4 | Conference on Advances in Communication and Control Systems | 3 | 3 |
| 1156 | Q4 | IEEE China Summit & International Conference on Signal and Information Processing (ChinaSIP) | 3 | 3 |
| 1156 | Q4 | IEEE International Conference on Emerging eLearning Technologies & Applications (ICETA) | 3 | 3 |
| 1156 | Q4 | International Conference on Actual Problems of Electronic Instrument Engineering | 3 | 3 |
| 1156 | Q4 | International Conference on Advanced Computer Science and Information Systems | 3 | 3 |
| 1156 | Q4 | International Conference on Circuits, Power and Computing Technologies | 3 | 3 |
| 1156 | Q4 | International Conference on e-Learning and e-Technologies in Education | 3 | 3 |
| 1156 | Q4 | International Conference on Information Communication and Embedded Systems | 3 | 3 |
| 1156 | Q4 | International Conference on Information Technology and Software Engineering | 3 | 3 |
| 1156 | Q4 | International Conference on Intelligent Systems and Control | 3 | 3 |
| 1156 | Q4 | International Convention on Information & Communication Technology Electronics & Microelectronics | 3 | 3 |
| 1169 | Q4 | International Conference on Electronics, Communications and Control | 2 | 52 |
| 1170 | Q4 | Nursing informatics...: proceedings of the... International Congress on Nursing Informatics | 2 | 5 |
| 1171 | Q4 | IET International Conference on Information Science and Control Engineering | 2 | 4 |
| 1171 | Q4 | International Conference on Advanced Computer Science and Electronics Information | 2 | 4 |
| 1171 | Q4 | International Conference on Engineering and Computer Education | 2 | 4 |
| 1171 | Q4 | International Conference Problems of Cybernetics and Informatics | 2 | 4 |
| 1171 | Q4 | International Symposium on Computer, Communication, Control and Automation | 2 | 4 |
| 1176 | Q4 | E3S Web of Conferences | 2 | 3 |
| 1176 | Q4 | IEEE International Scientific Conference Electronics and Nanotechnology (ELNANO) | 2 | 3 |
| 1176 | Q4 | International Conference on Electric Information and Control Engineering | 2 | 3 |
| 1176 | Q4 | International Conference on Electrical Engineering and Software Applications | 2 | 3 |
| 1176 | Q4 | International Conference on Electronic & Mechanical Engineering and Information Technology | 2 | 3 |
| 1176 | Q4 | International Conference on Future Computer Science and Education (ICFCSE) | 2 | 3 |
| 1176 | Q4 | International Conference on Optoelectronics and Microelectronics | 2 | 3 |
| 1176 | Q4 | International Conference on Thermal, Mechanical and Multi-Physics Simulation and Experiments in Microelectronics and Microsystems (EuroSimE) | 2 | 3 |
| 1176 | Q4 | Symposium of Image, Signal Processing, and Artificial Vision (STSIVA) | 2 | 3 |
| 1185 | Q4 | Asia-Pacific Signal and Information Processing Association Annual Summit and Conference (APSIPA) | 2 | 2 |
| 1185 | Q4 | IEEE International Conference on Oxide Materials for Electronic Engineering | 2 | 2 |
| 1185 | Q4 | International Conference on Control, Decision and Information Technologies | 2 | 2 |
| 1185 | Q4 | International Conference on Fuzzy Theory and Its Applications | 2 | 2 |
| 1185 | Q4 | International Congress on Advanced Electromagnetic Materials in Microwaves and Optics | 2 | 2 |
| 1185 | Q4 | International Journal of Wireless and Microwave Technologies (IJWMT) | 2 | 2 |
| 1185 | Q4 | International Workshop on Active-Matrix Flatpanel Displays and Devices | 2 | 2 |
| 1185 | Q4 | International Workshop on Microwave and Millimeter Wave Circuits and System Technology | 2 | 2 |
| 1193 | Q4 | European Conference on Information Technology in Education and Society: A Critical Insight | 1 | 3 |
| 1193 | Q4 | International Conference on Information Systems for Crisis Response and Management (ISCRAM) | 1 | 3 |
| 1195 | Q4 | International Conference on Actual Problems of Electron Devices Engineering | 1 | 2 |
| 1195 | Q4 | International Conference on Computing, Electrical and Electronics Engineering | 1 | 2 |
| 1195 | Q4 | International Conference on Electric and Electronics | 1 | 2 |
| 1195 | Q4 | International Conference on Information Security and Intelligence Control | 1 | 2 |
| 1195 | Q4 | International Conference on Information, Business and Education Technology | 1 | 2 |
| 1195 | Q4 | International Conference on Systems and Control | 1 | 2 |
| 1201 | Q4 | IEEE Global Conference on Signal and Information Processing (GlobalSIP) | 1 | 1 |
| 1201 | Q4 | International Future Energy Electronics Conference (IFEEC) | 1 | 1 |
| 1201 | Q4 | International Symposium on Computing and Networking | 1 | 1 |
| 1201 | Q4 | International Workshop on Cloud Computing and Information Security | 1 | 1 |
| 1201 | Q4 | Iranian Conference on Fuzzy Systems | 1 | 1 |
| 1201 | Q4 | The International Conference on Education Technology and Information System | 1 | 1 |
| 1201 | Q4 | Virtual Reality Society of Japan, Annual 日本 バーチャル リアリティ 学会 大会 論文 集 Conference | 1 | 1 |
| 1201 | Q4 | The International Conference on Remote Sensing, Environment and Transportation Engineering | 1 | 1 |

## ACKNOWLEDGEMENTS

This study has been funded under project HAR2011-30383-C02-02 from Dirección General de Investigación y Gestión del Plan Nacional de I+D+I (Ministry of Economy and Competitiveness), of which Emilio Delgado López-Cózar is the principal investigator, and project APOSTD/2013/002 from the Regional Ministry of Education, Culture and Sports (Generalitat Valenciana, Spain), awarded to Enrique Orduña-Malea. Alberto Martín-Martín enjoys a four-year doctoral fellowship (FPU2013/05863) granted by the Spanish Ministry of Education, Culture and Sports. Juan Manuel Ayllón enjoys a four-year doctoral fellowship (BES-2012-054980) granted by the Spanish Ministry of Economy and Competitiveness.